\documentclass[%
reprint,
 amsmath,amssymb,
 aps,
]{revtex4-2}
\usepackage{graphicx}
\usepackage{dcolumn}
\usepackage{bm}
\usepackage{braket}
\usepackage{graphicx}   

\usepackage{here}

\usepackage{color}

\begin{document}
\preprint{APS/123-QED}

\title{Study of decoherence in radial local phonon hopping
   within trapped-ion string}

\author{Yu-Xuan Chen$^1$}%
\email{u320161d@ecs.osaka-u.ac.jp}
\author{Takumi Yuri$^1$}
\author{Kenji Toyoda$^{2,1}$}%
\email{kenji.toyoda.qiqb@osaka-u.ac.jp}
\affiliation{$^1$Graduate School of Engineering Science, Osaka University, 1-3 Machikaneyama, Toyonaka, Osaka 560-8531, Japan}
\affiliation{$^2$Center for Quantum Information and Quantum Biology, Osaka University, 1-2 Machikaneyama, Toyonaka, Osaka 560-0043, Japan}

\date{\today}

\begin{abstract}
We systematically investigate local phonon hopping in the radial
direction of a linear trapped-ion string. We measure the decay of phonon
hopping as a function of key trap parameters and analyze the results in
terms of the decay time and the number of oscillations. We attribute the
loss of coherence to both nonlinear mode coupling and
environmental electric-potential noise.  Numerical simulations
incorporating these effects are performed and compared with the
experimental results. This work establishes a method for
evaluating phonon hopping coherence and provides insight
into decoherence mechanisms in radial local-phonon
dynamics.
\end{abstract}

\maketitle

\section{\label{sec:level1}Introduction}

A trapped-ion system is considered to be an excellent platform for quantum computing because of its exceptional performance in quantum information processing \cite{cirac1995quantum, blatt2008entangled}. It involves trapping ions using electromagnetic fields and applying laser pulses to precisely manipulate and measure quantum states \cite{leibfried2003quantum, haffner2008quantum}. This system has a long coherence time, high-fidelity quantum logic gate operation, and a highly scalable architecture, offering significant advantages for quantum computing, quantum simulations, and quantum sensing \cite{monroe2013scaling, brown2016co, porras2004effective, schmidt2003realization}.

In trapped-ion systems, phonons can be used as a medium for transferring
information between ions, whose internal states serve as qubits or
pseudospins.  By being associated with collective modes of motion,
phonons enable the coupling and entanglement of quantum states through
Coulomb interactions between ions and have a large information capacity
due to the high dimensionality of their Hilbert space.
Accurate manipulation of phonons is crucial for achieving precise and fast quantum computations, especially for realizing multi-ion quantum state entanglement and quantum gate operations.
In addition, phonons can be used to simulate physical models that involve bosonic particles, such as the Bose-Hubbard model \cite{zwerger2003mott, greiner2002quantum}, the spin-boson model \cite{leggett1987dynamics}, and the Holstein model \cite{holstein1959studies, fehske2007numerical}.
The vibrational modes in trapped ions and phonons have been demonstrated to be fundamental quantum computing building blocks  \cite{Chen2023,Gan2020,Chen2021}.

Phonons in trapped ions can be categorized as either collective-mode phonons or local phonons. 
For collective-mode phonons, ions collectively oscillate due to strong Coulomb interactions
\cite{james1998quantum,Zhu2006}.
For local phonons, ions oscillate in modes that are approximately independent of each other because the confinement force acting on each ion in one direction is much stronger than the Coulomb forces. The ions can still be weakly coupled together by Coulomb interactions.

Studies of local phonons in trapped ions can be divided into those of
phonons in multi-well potentials or trap arrays and those of phonons in
the radial direction of a linear ion string.  For the former type of
phonon, each site can be controlled independently.  Coupled quantized
mechanical oscillators and the exchange (hopping) of local phonons
between them have been examined using two ions in double-well potentials
\cite{Brown2011,Harlander2011}.  Multiple potential wells, each
containing a single ion, have been controlled to demonstrate the tunable
coupling of phonons \cite{Wilson2014}, their coherent coupling
\cite{Hou2024}, and two-dimensional networks of vibrational modes
\cite{Hakelberg2019,Kiefer2019}.  Regarding the latter type of phonon,
Porras and Cirac proposed using local phonons in the radial direction of
a linear ion string for simulating the Bose-Hubbard model
\cite{porras2004bose}.  The hopping of radial local phonons and their
interference have been investigated
\cite{haze2012observation,Debnath2018,Tamura2020,Toyoda2015}.  Radial
local phonons have been used to simulate the Jaynes-Cummings-Hubbard
model and the Rabi-Hubbard model
\cite{Ivanov2009,Toyoda2013,Li2022,Mei2022}.

A system of
local phonons in the radial direction of a linear ion string can be
prepared in a standard linear trap 
and applied to quantum simulations or analog quantum computation \cite{Shen2014}.
Despite the importance of radial local phonons,
some of their basic characteristics that may affect experimental implementations are not fully understood.
One of the most important of these characteristics is the coherence
in the phonon-hopping process.
The hopping of phonons occurs as an elementary process 
that can be expressed using a hopping Hamiltonian
\cite{porras2004bose},
which is key to applications using radial phonons.
The process and coherence may be affected by various factors,
including the residual thermal distribution in the system.
The conditions for a well-defined concept of radial phonons, 
especially their number preservation,
require that the confinement force be much stronger than Coulomb forces, as mentioned above.
To satisfy this condition,
the axial confinement should be kept weak
so that the inter-ion distances become relatively large
(in our typical conditions with $^{40}$Ca$^+$ ions, we use 
inter-ion distances of $\sim$20 $\mu$m 
for a radial trap frequency of $\sim$3 MHz).
Due to the relatively weak confinement along the axial direction
and the resulting lack of Lamb-Dicke confinement,
it is not straightforward 
to apply sub-Doppler or ground-state cooling mechanisms
such as sideband cooling.
We can avoid the direct effect of the axial thermal distribution 
(e.g., the appearance of Doppler sidebands)
by setting the direction of the excitation lasers
for the manipulation and observation of radial phonons
perpendicular to the trap axis.
The indirect effects of the residual axial thermal distribution,
especially those on the coherence of phonon hopping,
have not been previously investigated.

In this study, we investigate the hopping process of radial
local phonons in a trapped-ion system, focusing on the coherence
preservation and decoherence mechanisms for phonons in localized
regions. We analyze phonon-hopping coherence in terms of decay time and
the number of oscillations.
Regarding the decoherence mechanism, we consider multiple
physical processes that can affect phonon-hopping coherence.
The Coulomb interaction between the ions induces a nonlinear coupling
between the vibrational modes in the $y$ and $z$ directions, giving rise
to a Kerr-type Hamiltonian.
In addition, environmental electric-potential noise is another
relevant source of decoherence in radial phonon dynamics. We investigate
the effects of these factors and compare their contributions to the
experimentally observed decay in the phonon-hopping
process.


\section{Experimental procedure}

We confined two $^{40}$Ca$^+$ ions in a linear Paul trap, with
radio-frequency (RF) and direct-current fields providing radial and
axial confinement. RF signals were fed to the trap electrodes through a
helical resonator with a resonance frequency of 23.5 MHz. Regulating the
RF amplitude of the radial electrode voltage allowed us to stabilize
the radial trap frequencies ($\omega _x$, $\omega _y$)
\cite{johnson2016active}. The confinement along the axial direction was
relatively weak, resulting in large inter-ion distances and the
formation of a linear ion chain in the axial direction.  We decomposed
the motion of the ions into three-dimensional vibrations in axial ($z$)
and radial ($x$ and $y$) directions.  The Hamiltonian governing the
radial direction (the $y$ direction in this study) can be written as
follows \cite{porras2004bose}:
\begin{equation}
 \label{eqn-Hy1}
 \hat{H}_y=
  \sum_{i=1,2}^{}{\frac{\hat{p}_{i}^{2}}{2m}}
  +\sum_{i=1,2}^{}{\frac{1}{2}m\omega_{y}^{2}\hat{y}_{i}^{2}}
  -\frac{q^2}
  {8\pi \varepsilon_0}
  \frac{(\hat{y}_1-\hat{y}_2)^2}{|\bar{z}_1-\bar{z}_2|^3},    
\end{equation}
where $\hat{y}_i$ and $\hat{p}_i$ represent the position and momentum, respectively, of
the $i$th ion in the ion chain along the radial $y$ direction, $m$ is the ion mass, $\omega_y$ is the trap frequency in
the radial $y$ direction, $q$ is the charge of each ion, and $\bar{z}_i$
represents the equilibrium position of the $i$th ion along the axial $z$
direction. The first two terms in the Hamiltonian represent the kinetic
energy and trap potential, respectively, and the last term represents
the Coulomb interaction, which induces phonon hopping.  We converted
the Hamiltonian in Eq.~\eqref{eqn-Hy1} to its second quantized form. It
can be rewritten using local-phonon operators as follows
\cite{porras2004bose}:
\begin{equation}
\label{eqn-Hy2}
    \hat{H}_y
    =\sum_{i=1,2}^{} \hbar 
    \left( \omega _y-\frac{\kappa}{2} \right) 
    \hat{a}_i^\dagger \hat{a}_i
    +\frac{\hbar\kappa}{2}
    \left(
     \hat{a}_1 \hat{a}_{2}^\dagger
     +\hat{a}_1^\dagger\hat{a}_2
    \right), 
\end{equation}
where
\begin{equation}
 \label{eqn-kappa}
  \kappa =\frac{q^2}{4\pi\varepsilon_0m\omega_yd_0^3}
\end{equation}
is the hopping rate between ions 1 and 2.  $\hat{a}_i$ and
$\hat{a}^\dagger_i$ are annihilation and creation operators for the
local phonon modes along the $y$ direction of the $i$th ion,
respectively. $d_0$ is the inter-ion distance.  Phonon hopping can also
be explained as the normal mode of radial vibrational motion. Our
previous study describes the specific details
\cite{haze2012observation}.

Two laser beams with wavelengths of 423 and 375 nm
were utilized for optical ionization. A two-step laser cooling technique was used to cool the ions to near the vibrational ground state. This consisted of Doppler cooling using the 397-nm ($S_{1/2}$--$P_{1/2}$) and 866-nm ($D_{3/2}$--$P_{1/2}$) transitions and resolved sideband cooling using 
the 729-nm ($S_{1/2}$--$D_{5/2}$)
and 854-nm ($D_{5/2}$--$P_{3/2}$) transitions. 
After the resolved sideband cooling, the average phonon numbers in the radial directions were $\left( \left< n_x \right> ,\left< n_y \right> \right)  \sim (0.30,0.04)$. The output of the 729-nm laser was split into two beams, each of which illuminated one of the two ions. Each beam was controlled by a dedicated single-pass acousto-optic modulator. This system of single-pass acousto-optic modulators allowed us to maintain the intensity of the laser beams irradiating the ions and 
equalizing the Rabi frequencies 
between the ions. The maximum carrier Rabi frequency for the $S_{1/2}(m_J=-1/2)$--$D_{5/2}(m_J=-1/2)$ transition was $\sim$800 kHz and the corresponding Rabi frequencies for the sideband transitions involving the vibrational ground state were $\sim$35 kHz.

\begin{figure}
    \centering
    \includegraphics[width=86mm]{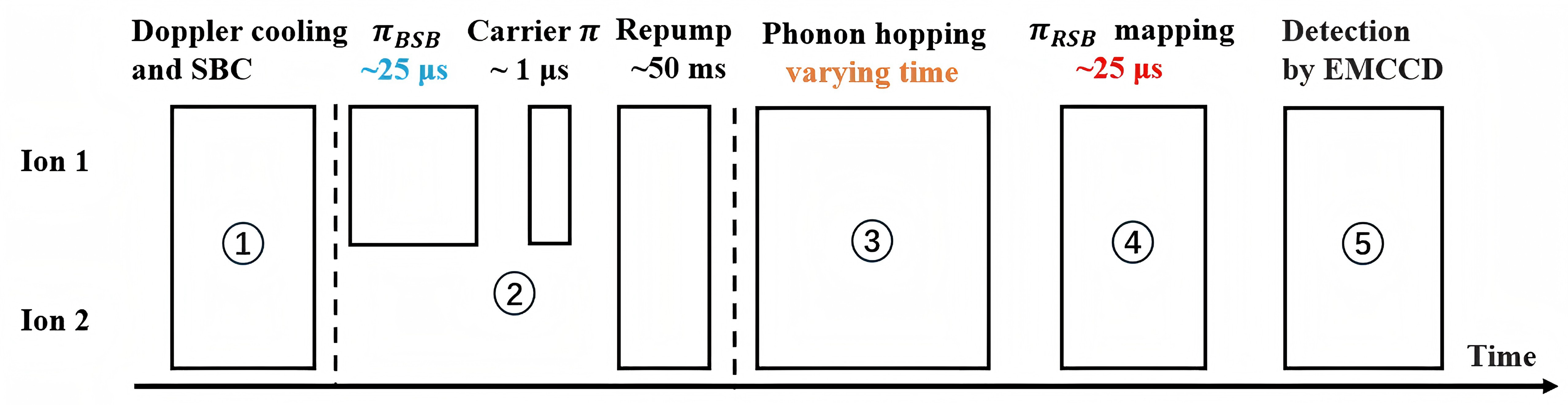}
    \caption{Experimental time sequence for observing phonon hopping. SBC: sideband cooling; RSB/BSB: red/blue sideband; EMCCD: electron-multiplying charge-coupled device. The circled numbers correspond to the steps in the sequence outlined in the main text.}
    \label{fig:time-sequence}
\end{figure}

The specific steps for observing phonon hopping are as follows (see Fig.~\ref{fig:time-sequence} for the experimental time sequence). 
\begin{enumerate}
 \item 
Perform Doppler and sideband cooling to cool the ions to their vibrational ground state. 
 \item 
Apply blue-sideband and carrier $\pi$ pulses to one of the two ions (referred to as ion 1) to prepare the initial phonon state $|1\rangle _1|0\rangle _2$. 
Immediately after, apply an 854-nm laser pulse near resonance with the $D_{5/2}-P_{3/2}$  transition to pump the residual population in
$D_{5/2}$. 
 \item 
Turn off all the lasers and let the phonon undergo hopping between the ions due to Coulomb interactions. 
 \item 
Irradiate both ions with a red-sideband $\pi$ pulse to map the vibrational states onto the internal states. Applying a red-sideband $\pi$ pulse causes an ion with a phonon number of 1 to be transferred to the internal excited state. No transition occurs if the phonon number is 0 (i.e., the ground state is maintained). Determine the internal state using the shelving method. 
 \item 
Illuminate both ions with a 397-nm laser and use an electron-multiplying charge-coupled-device camera to image the fluorescence to detect the internal state of individual ions.
\end{enumerate}

\section{Measurement of phonon-hopping dynamics}
According to Eq.~\eqref{eqn-kappa}, the hopping rate depends on 
$d_0$, the inter-ion distance (which depends on the axial confinement),
and $\omega_y$, the trap frequency in the $y$ direction.
These parameters were varied in our systematic investigation of phonon-hopping decay. 

In typical ion trap experiments, including the present experiment, the parameters given
above are used to control the binding conditions of ions in the
trap.
As a result, they determine 
the configuration and motion of the ions.
As discussed later, the factors that may affect
the coherence of phonon hopping, such as nonlinear couplings between 
collective modes, may depend on factors such as the strength of inter-mode
couplings and the velocity distribution in thermal states.
These factors largely depend on the configuration and motion of ions 
and hence on the global trap parameters, as
stated above.
We thus chose the above parameters for the systematic investigation.

To quantify the coherence of phonon-hopping dynamics, 
we introduce the {\it number of oscillations}, 
defined as 
the number of cycles 
in the phonon-hopping process from the initial time until the contrast decreases to $1/e$ of its initial value.

\subsection{Results for phonon-hopping dynamics}
Figures 2 and 3 show data corresponding to the cases of the maximum
number of oscillations and longest decay time in our experiment,
respectively. The blue points represent the probability of being
detected in an excited state, with each data point averaged over 20
measurements. The hopping rate is determined by fitting a sinusoidal
function. The obtained values are generally consistent with the
theoretical prediction from Eq.~\eqref{eqn-kappa}. This analysis
confirms the expected behavior of phonon-hopping dynamics under various
trap conditions, which is consistent with our theoretical model.

\begin{figure} 
    \centering \includegraphics[width=86mm]{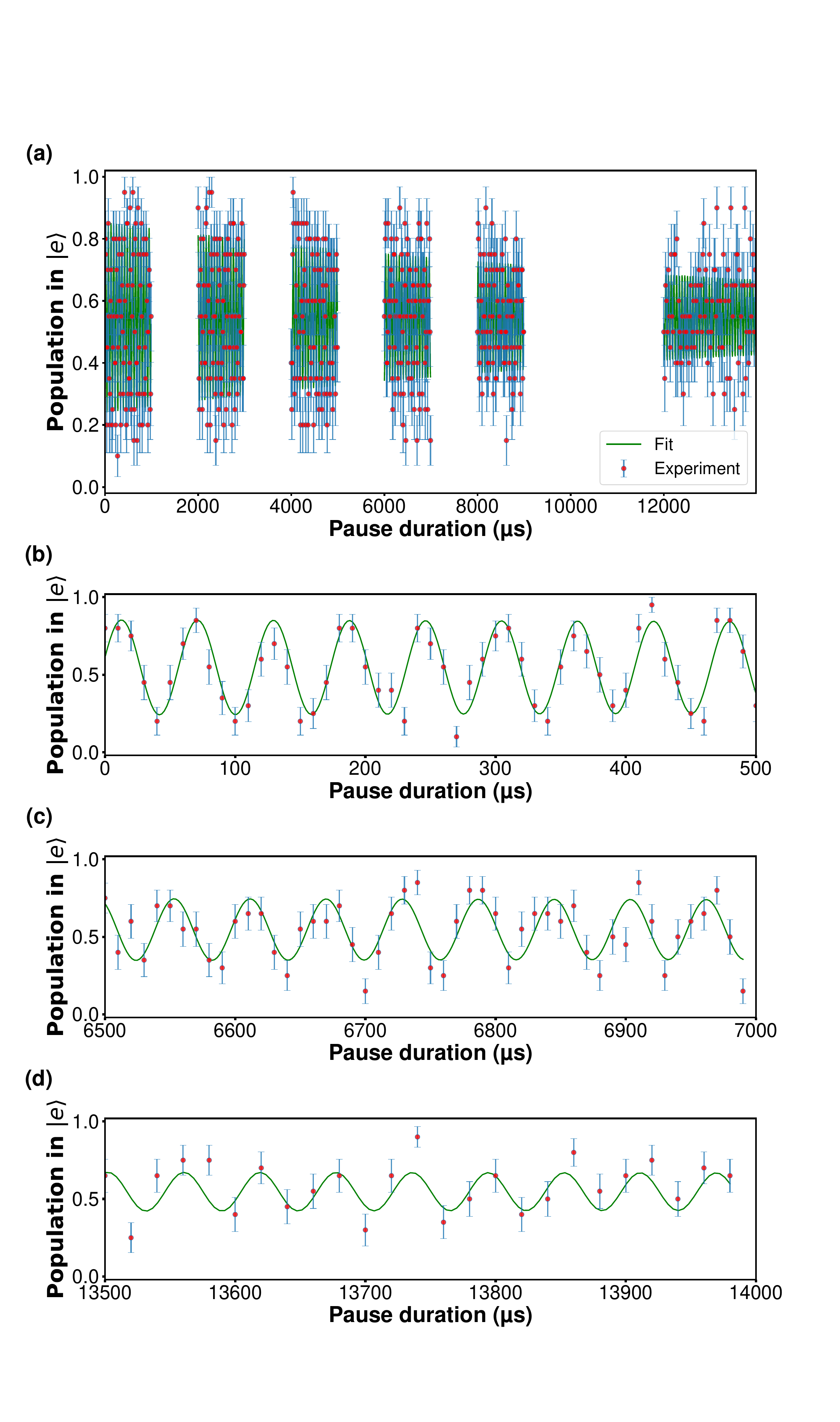}
    \caption{Phonon-hopping results corresponding to case of the
    experimentally obtained maximum number of oscillations ($268.3 \pm
    25.8$) in our experiment.  The corresponding radial trap frequency
    $\omega_y/2\pi$ is 2.85 MHz and the inter-ion distance $d_0$ is 12.2
    $\mu$m. The blue dots are the experimental measurements and the
    green curve is the result obtained from fitting the function
    $ae^{-bx}\sin (cx+d) +fx$. Each point in the
    experimental results is the average of 20 measurements, and the
    error bars represent the statistical uncertainty of the measured
    population arising from the finite number of experimental
    repetitions. (a) Full experimental data and numerical
    calculations for phonon hopping. (b-d) Magnified views of different
    time periods for phonon-hopping process.}  \label{fig:hopping1}
\end{figure}

\begin{figure} 
    \centering \includegraphics[width=86mm]{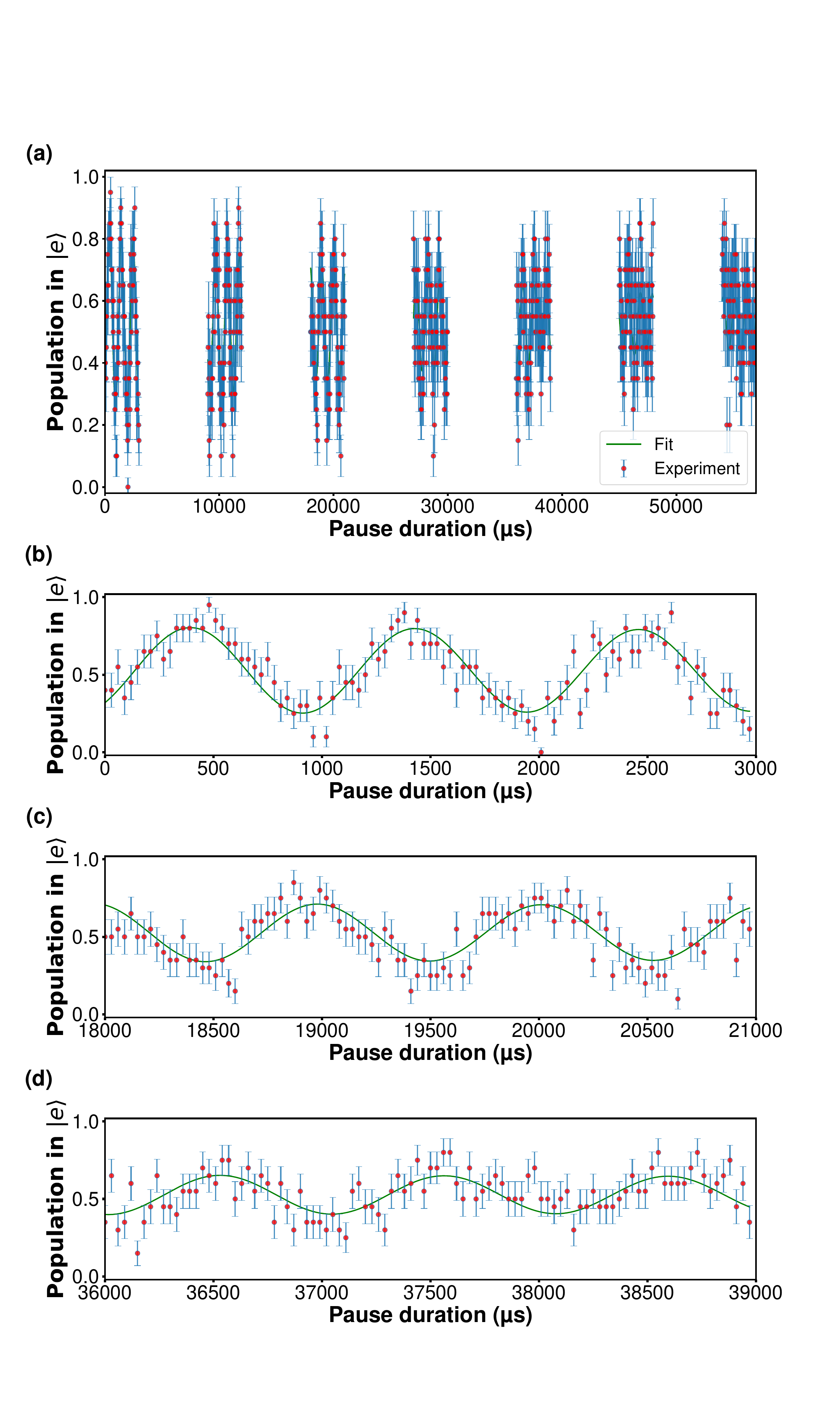}
    \caption{Phonon-hopping results corresponding to case of longest
    decay time in our experiment ($42.5 \pm 3.3~\mathrm{ms}$).  The
    corresponding radial trap frequency $\omega_y/2\pi$ is 2.85 MHz and
    the inter-ion distance $d_0$ is 31.7 $\mu$m.  The blue dots are the
    experimental measurements and the green curve is the result obtained
    from fitting the function $ae^{-bx}\sin (cx+d) +fx$. 
    Each point in the experimental results is the average of 20
    measurements, and the error bars represent the statistical
    uncertainty of the measured population arising from the finite
    number of experimental repetitions. (a) Full
    experimental data and numerical calculations for phonon
    hopping. (b-d) Magnified views of different time periods for
    phonon-hopping process.}  \label{fig:hopping2}
\end{figure}

\subsection{Systematic investigation of dependence on trap parameters}
To investigate the dependence of the phonon-hopping coherence on $d_0$,
we controlled $d_0$ by varying the axial electrode voltage while
maintaining a fixed value of 2.85 MHz for $\omega_y/2\pi$. The relation
between $d_0$ and the axial trap frequency $\omega_z$ assumed here is
$d_0 = \left( e^2 / 4\pi \varepsilon_0 m \omega_z^2 \right)^{1/3} \left(
2.018 / 2^{0.559} \right) $ \cite{james1998quantum}.  The values of
$d_0$ examined here are $\{12.2,15.9,19.4,31.7\}$ $\mu$m, which
correspond to $\omega_z/2\pi$ values of $\{313,210,155,74\}$ kHz,
respectively.

To investigate the dependence of the phonon-hopping coherence on the
radial trap frequency $\omega_y$, we fixed the inter-ion distance $d_0$
at 19.1 $\mu$m and varied $\omega_y$ by adjusting the RF amplitude of
the radial electrode voltage.  The values of $\omega_y/2\pi$ examined
here are $\{2.43, 2.64, 2.85, 3.11\}$ MHz.

The experimental results in 
Figs.~4 and 5 
show the dependence of phonon-hopping coherence on the inter-ion distance $d_0$ and the radial trap frequency $\omega_y$. 
In Fig.~4, the two plots show the decay time on the vertical axis, while Fig. 5 shows the number of oscillations.
In both figures, 
the red circles with error bars are the results of the experiment. (The numerical simulations shown in Figs. 4 and 5 are discussed in a following section.)

\begin{figure}
    \centering \includegraphics[width=86mm]{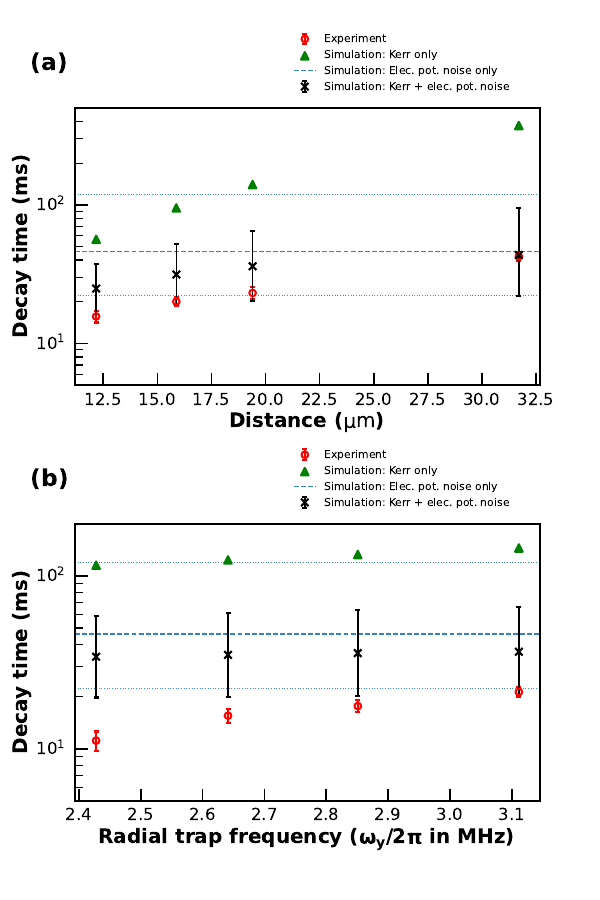}
    \caption{Experimental results for decay time
    (red circles with error bars) compared with numerical simulations
    based on the Kerr-only model (green
    triangles) and a combined Kerr and electric-potential noise model  (black
    crosses).  Numerically simulated results in the case of only
    electric-potential noise are also indicated by blue dashed curves
    (central value) and blue dotted curves (lower and upper bounds of
    the confidence interval). (a) Decay time against 
    inter-ion distance $d_0 =
    \{12.2,\,15.9,\,19.4,\,31.7\}\,\mu\mathrm{m}$ with the radial trap
    frequency fixed at $\omega_y/2\pi = 2.85\,\mathrm{MHz}$.  (b) Decay
    time against radial trap frequency $\omega_y/2\pi =
    \{2.43,\,2.64,\,2.85,\,3.11\}\,\mathrm{MHz}$ with the inter-ion
    distance fixed at $d_0 = 19.1\,\mu\mathrm{m}$.
    }\label{fig:decay-time}
\end{figure}

\begin{figure}
    \centering \includegraphics[width=86mm]{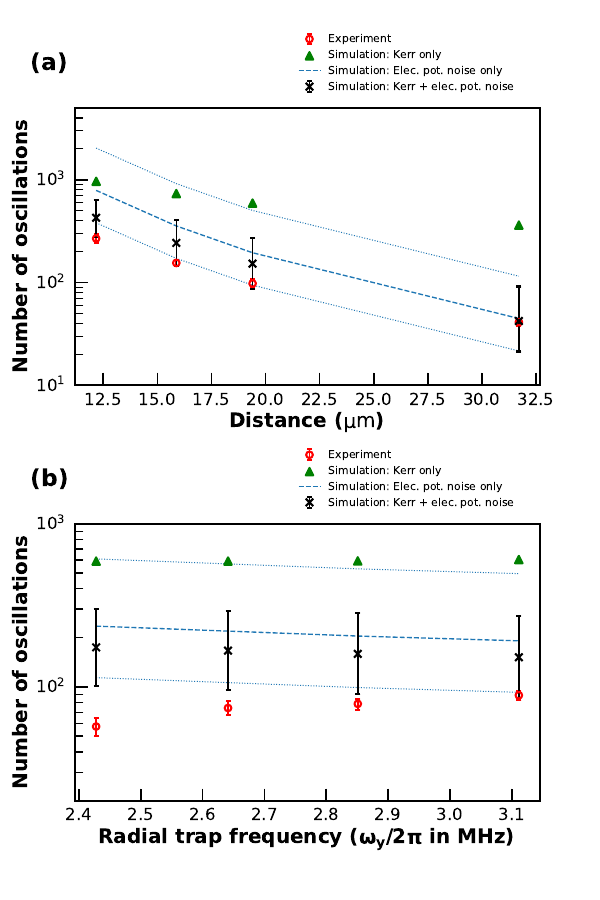}
    \caption{Experimental results for  
    number of oscillations (red circles with error bars) compared with
    numerical simulations based on the Kerr-only
    model (green triangles) and a combined Kerr and electric-potential noise model (black crosses).  
    Numerically simulated results in the case of
    only electric-potential noise are also indicated by blue dashed
    curves (central value) and blue dotted curves (lower and upper
    bounds of the confidence interval). (a) Number of
    oscillations against inter-ion distance $d_0 =
    \{12.2,\,15.9,\,19.4,\,31.7\}\,\mu\mathrm{m}$ with the radial trap
    frequency fixed at $\omega_y/2\pi = 2.85\,\mathrm{MHz}$.  (b) Number
    of oscillations against radial trap frequency $\omega_y/2\pi =
    \{2.43,\,2.64,\,2.85,\,3.11\}\,\mathrm{MHz}$ with the inter-ion
    distance fixed at $d_0 = 19.1\,\mu\mathrm{m}$.  }
    \label{fig:number-of-oscillations}
\end{figure}

As shown in Fig.~4(a), the decay time significantly increases with inter-ion distance, indicating that phonon coherence is better maintained at larger distances. 
Similarly, Fig.~4(b) shows that the decay time increases with increasing trap frequency. 
Fig.~5(a) shows that the number of oscillations decreases with increasing inter-ion distance.
Conversely, Fig.~5(b) shows that the number of oscillations increases with increasing trap frequencies, which suggests enhanced coherence. 
In the next section, we introduce a model based on a nonlinear coupling
between collective modes and interpret the global trends shown in
Figs.~4 and 5.

\section{Effect of inter-mode couplings due to Kerr nonlinearity on phonon-hopping decoherence}
\label{sec:inter-mode-couplings}

In our experiment, the ions were cooled to the vibrational ground state
in the $x$ and $y$ directions by sideband cooling. On the other hand,
the condition of Lamb-Dicke confinement, which is almost a prerequisite
for performing sideband cooling, was not satisfied in the $z$ direction;
only Doppler cooling was performed in this direction.  For the values of
axial trap frequency given above, the modified Lamb-Dicke parameter
\(\eta \sqrt{n_{\text{av}}}\), which accounts for finite temperature
effects, ranges from 1.0 to 4.4.  This indicates that the system was not
fully within the Lamb-Dicke regime, as \(\eta \sqrt{n_{\text{av}}} < 1\)
was not satisfied for the given range.  In the case of two ions, the
motion of the ions in the $z$ direction can be decomposed into the
center-of-mass (COM) and stretching modes. In the stretching mode, the
ions oscillate with opposite phases, which leads to an inter-ion
distance change. It is known that the stretching mode can be
preferentially coupled to the rocking mode in the $y$ direction, which
also involves an inter-ion distance change via the nonlinearity in the
Coulomb interactions \cite{roos2008nonlinear,nie2009theory}.  Here, we
analyze the effect of this nonlinear mode coupling on phonon-hopping
decoherence.

\subsection{Model of nonlinear mode coupling between radial and axial modes}
The inter-ion distance $d_0$ influences the Coulomb interactions between
ions.  By varying $d_0$, we modified the strength of these interactions,
directly influencing vibrational mode coupling. A larger $d_0$ reduces
the Coulomb interaction strength, leading to weaker mode coupling. This
reduction in coupling lowers the nonlinear coefficients in the
phonon-phonon coupling Hamiltonian, such as the Kerr-type Hamiltonian:
\begin{equation}
 \label{eqn-HKerr}
  \hat{H}_{\text{Kerr}}=\hbar \chi \hat{n}_r\hat{n}_s,
\end{equation}
where $\hat{n}_r$ and $\hat{n}_s$ are 
the phonon number operators for the rocking mode in the $y$ direction
and the stretching mode in the $z$ direction, respectively. The coupling
constant $\chi$ is expressed as \cite{nie2009theory}:

\begin{equation}
\label{eqn-chi}
\chi =-\omega _s\left( \frac{1}{2}+\frac{\omega _{s}^{2}/2}{4\omega _{r}^{2}-\omega _{s}^{2}} \right) \left( \frac{\omega _z}{\omega _r} \right) \left( \frac{2\hbar \omega _z}{\alpha ^2mc^2} \right) ^{1/3},
\end{equation}
where $\alpha$ denotes the fine-structure constant, $c$ is the speed of light, and $\omega _s$ and $\omega _r$ are the stretching- and rocking-mode frequencies,  respectively. 
Similarly, the radial trap frequency $\omega_y$ defines the ion confinement strength in the radial direction. By varying $\omega_y$, we altered the vibrational states and energy levels, significantly impacting phonon dynamics. 

The Hamiltonian for the Kerr-type mode coupling in Eq.~\eqref{eqn-HKerr}
can be interpreted 
in the collective-mode Fock-state basis
as a shift of the rocking mode frequency, which is proportional to the stretching-mode quantum number $n_s$, as follows:
\begin{equation}
 \label{eqn-delta-omega-r}
  \delta \omega _r=\chi n_s.
\end{equation}

\subsection{Principles of numerical simulation incorporating inter-mode couplings}
\label{sec:principles-inter-mode-couplings}
To evaluate the magnitude of the shift of the rocking mode using 
Eqs.~\eqref{eqn-chi} and \eqref{eqn-delta-omega-r}, 
we need the trap frequencies and the average vibrational quantum numbers along the axial direction. The axial trap frequency in various conditions can be evaluated with axial sideband spectroscopy in combination with inter-ion distance measurements using fluorescence images. To evaluate the average vibrational quantum numbers along the axial direction, the heights of the axial sidebands can be used. They can be compared with a model for which a thermal quantum number distribution is assumed.

Since the axial motion was not in the Lamb-Dicke regime in our typical experimental conditions and we observed many sideband spectra, we did not fit model functions to each spectral component. Instead, we used a single Doppler-broadened envelope to fit the entire spectrum. We assumed that the axial motion is in thermal equilibrium with a temperature close to the Doppler limit $T_D=\hbar \varGamma /2k_B$. In the case of $^{40}$Ca$^+$ ions, the natural width of the $S_{1/2}-P_{1/2}$ transition is 20.4 MHz \cite{cui2022evaluation,Huang2014Evaluation}. The Doppler-limited temperature calculated on this basis is 490 $\mu$K. The Hamiltonian for the harmonic motion of the ions in the $z$ direction can be described as follows:
\begin{equation}
 \label{eqn-Hz}
  \hat{H}_z=\sum_{i=1,2}
  \left(
   \frac{\hat{p}_{z,k}^2}{2m} + \frac{1}{2}m\omega_z^2\hat{z}_k^2 
   +\hat{V}_{\mathrm{Coul},z}
  \right), 
\end{equation}
where \(\hat{p}_{z,i}\) is the momentum of the \(i\)th ion along the
\(z\) axis, \(\hat{z}_i\) is the position coordinate of the \(i\)th
ion, \(m\) is the mass of the ion, \(\omega_z\) is the trap
frequency in the \(z\) direction, and $\hat{V}_{\mathrm{Coul},z}$ is the
Coulomb energy part of the potential energy in the $z$ direction.

According to the equipartition theorem in thermodynamics
\cite{reif2009fundamentals}, the average kinetic energy per degree of
freedom at temperature \(T\) is \(\frac{1}{2}k_B T\).  We assumed that
both ions are in thermal equilibrium.  The sum of the average kinetic
energies is given by:
\begin{equation}
  \left. \sum_{i=1,2}{\frac{\langle{}\hat{p}_{z,i}^{2}\rangle}{2m}} \right. =k_BT,
\end{equation}
where $\langle\cdots\rangle$ represents an ensemble average with respect to the
thermal distribution.  This can be rewritten using
$\hat{v}_{z,i}\equiv\hat{p}_{z,i}/m$ ($i=1,2$), the velocity of the
$i$th ion:
\begin{equation}
  \left\langle \frac{1}{2} m\hat{v}_{z,1}^2 + \frac{1}{2} m\hat{v}_{z,2}^2 \right\rangle = k_B T.
\end{equation}
Assuming that the mean square velocities of both ions are equal,
\(\left\langle \hat{v}_{z,1}^2 \right\rangle = \left\langle \hat{v}_{z,2}^2
\right\rangle\), we obtain the following relation for the root mean
square (rms) velocity $v_{\text{rms}} = \sqrt{\langle \hat{v}_{z,i}^2\rangle}$
($i=1,2$):
\begin{equation}
 \label{eqn-m-vrms2}
  mv_{\text{rms}}^2 = k_B T.
\end{equation}
From the equipartition theorem, we can also obtain the average total energy for the axial stretching mode. 
It can be represented as follows, considering that a harmonic oscillator mode has both kinetic and potential energies:
\begin{equation}
 \label{eqn-hbar-omegas-ns}
  \hbar\omega_s\left\langle\hat{n}_s\right\rangle=k_BT,
\end{equation}
where $\omega_s$ and $\langle\hat{n}_s\rangle$ are the oscillation
frequency and the average quantum number for the axial stretching mode,
respectively.  From Eqs.~\eqref{eqn-m-vrms2} and
\eqref{eqn-hbar-omegas-ns}, we can relate $\langle\hat{n}_s\rangle$ and
$v_{\text{rms}}$ as follows:
\begin{equation}
 \label{ns}
  \langle\hat{n}_s\rangle=\frac{mv_{\text{rms}}^2}{\hbar\omega_s}.
\end{equation}

The trap frequency and the average vibrational quantum number are obtained using the relation 
$d_0$=$({e^2}/{4\pi \varepsilon _0m\omega _{z}^{2}})({2.018}/{2^{0.559}})$ \cite{james1998quantum}
and Eq.~\eqref{eqn-Hz}, respectively. 
The expression for calculating the phonon-hopping decay rate is as follows:

\begin{equation}
 h_\mathrm{Kerr}\left( t \right) =\sum_{n_s=0}^N{P(n_s) \sin ^2\left[ \frac{\left( \kappa -\chi n_s \right) t}{2} \right]},
  \label{eqn-hKerr}
\end{equation}
where $P({n_s})$ is the phonon number distribution for the axial stretching
mode, which is calculated by assuming a thermal distribution based on
Eq.~(\ref{ns}) with $v_\text{rms}$ determined from experimental
Doppler-broadened spectra.  

\section{Effect of electric-potential noise on phonon-hopping decoherence}
\label{sec:elec-pot-noise}

In the previous section, we showed that Coulomb-induced
Kerr-type nonlinear coupling between different vibrational modes
provides a natural mechanism for coherence loss in phonon-hopping
dynamics.   
In this section, we consider the effect of
electric-potential noise on phonon-hopping coherence another important
mechanism for coherence loss.

\subsection{Dephasing due to electric-potential noise}
\label{sec:dephasing} In trapped-ion systems, it is widely recognized
that electric-potential noise from the surrounding environment
is an important source of imperfections affecting the motional
states.  Such noise and its effects are most commonly discussed in the
context of motional heating \cite{turchette2000heating}.  In this
context, fluctuating electric fields (the first derivatives of
fluctuating electric potentials), which have frequency components around
the motional frequencies of the trapped-ion string, cause heating of
motional modes.  Fluctuating electric potentials, especially their
higher-order derivatives (e.g., the second-order derivative, equivalent
to potential curvature or electric-field gradient), can also cause dephasing, a
different type of imperfection.  In this subsection,
we investigate dephasing due to electric-potential noise, which leads
to phonon-hopping decoherence.

We first review the formalism for the dynamics in a radial vibrational mode
(Sec.~\ref{sec:formalism-for-radial-mode}) and confirm that a radial
local-phonon state can be represented as the superposition of radial
collective-mode states (COM and rocking modes in the case of two ions).  

Next, in
Sec.~\ref{sec:dephasing-2} we introduce fluctuating phase factors and
thereby describe the explicit form of a density-matrix element
relevant to the superposition state and hence to phonon-hopping
coherence [Eq.~\eqref{eqn-COM-rho2-Rock}].  

We then consider the effect
of noise on the phase factors (Sec.~\ref{sec:dephasing-3}).  We
introduce a noise spectrum and a corresponding decay rate
[$S_{\Delta\omega}(\omega)$ and $\Gamma_{\Delta\omega}$ in
Eq.~\eqref{eqn-S-Delta-omega}], and confirm that the average of a
fluctuating phase factor can be represented as an exponentially decaying
factor whose decay rate is proportional to a noise-spectral component.
We can therefore connect phonon-hopping decoherence to the noise
spectral component.

\subsubsection{Formalism for dynamics in radial vibrational mode}
\label{sec:formalism-for-radial-mode}
We assume a system of two ions confined in a linear trap, similar to
that discussed previously.  Here, the Hamiltonian in
Eq.~\eqref{eqn-Hy2} is slightly modified to incorporate site-dependent
trap frequencies: $\omega_{y,i}=\omega_y+\delta\omega_{y,i}$ ($i=1,2$).
$\delta\omega_{y,i}=\delta\omega_{y,i}(t)$ represents the site-dependent
shifts of the trap frequency due to the fluctuating
second-order derivative of the electric potential.

We assume that only one phonon is present in the system, for which
case the Hamiltonian can be represented as a matrix instead of an
algebraic combination of operators as in Eq.~\eqref{eqn-Hy2}.  Using an interaction picture with respect to the rotating frame
oscillating at frequency $\omega_y$, the Hamiltonian is represented as:
\begin{equation}
\label{eqn-Hamiltonian-HyI}
H_{y\mathrm{I}}=\hbar
\begin{pmatrix}
\delta\omega_{y,1}-{\kappa}/{2} & {\kappa}/{2} \\
{\kappa}/{2} & \delta\omega_{y,2}-{\kappa}/{2} \\
\end{pmatrix},
\end{equation}
where the basis for the matrix is represented using two-site motional Fock states
as $\{\ket{1,0},\ket{0,1}\}$.

We first neglect the fluctuating frequency shifts
($\delta\omega_{y,i}=0$ for $i=1,2$) and obtain the eigenstates.
Corresponding to the two eigenvectors,
$\mathbf{x}_\mathrm{COM}=(1/\sqrt{2})(1,1)$ and
$\mathbf{x}_\mathrm{Rock}=(1/\sqrt{2})(1,-1)$, two eigenstates are
obtained:
\begin{eqnarray}
 \ket{\mathrm{COM}(0)}&=&\frac{1}{\sqrt{2}}(\ket{1,0}+\ket{0,1}),\\
 \ket{\mathrm{Rock}(0)}&=&\frac{1}{\sqrt{2}}(\ket{1,0}-\ket{0,1}),
\end{eqnarray}
and the time-dependent state kets are:
\begin{eqnarray}
 \ket{\mathrm{COM}(t)}&=&\frac{1}{\sqrt{2}}e^{{-i\kappa{}t}/{2}}(\ket{1,0}+\ket{0,1}),
\label{eqn-COMt}\\
 \ket{\mathrm{Rock}(t)}&=&\frac{1}{\sqrt{2}}e^{{i\kappa{}t}/{2}}(\ket{1,0}-\ket{0,1}).
\label{eqn-Rockt}
\end{eqnarray}
These are states with one phonon in either of the two radial collective modes,
namely the COM and rocking modes.

We can show that local-phonon states can be represented as superpositions
of those collective-mode states:
\begin{eqnarray}
 \ket{1,0}&=&\frac{1}{\sqrt{2}}(\ket{\mathrm{COM}(0)}+\ket{\mathrm{Rock}(0)}),\\
 \ket{0,1}&=&\frac{1}{\sqrt{2}}(\ket{\mathrm{COM}(0)}-\ket{\mathrm{Rock}(0)}).
\end{eqnarray}
If we prepare the local-phonon state $\ket{1,0}$ at $t=0$, the following
time-dependence is expected:
\begin{equation}
 \ket{\psi_1(t)}=\frac{1}{\sqrt{2}}(\ket{\mathrm{COM}(t)}+\ket{\mathrm{Rock}(t)}).
\label{eqn-psi1}
\end{equation}
Thus, the coherence of phonon hopping induced by the Hamiltonian in
Eq.~\eqref{eqn-Hamiltonian-HyI} is equivalent to the
coherence between $\ket{\mathrm{COM}(t)}$ and $\ket{\mathrm{Rock}(t)}$,
which constitute the superposition state.

\subsubsection{Fluctuating phase factors affecting phonon-hopping coherence}
\label{sec:dephasing-2}

Next, we consider the effect of fluctuating phase factors due to the stochastic process represented by the variables
$\delta\omega_{y,i}$ ($i=1,2$), by assuming that they are non-zero fluctuating quantities.
The state under the influence of the stochastic process, $\ket{\psi_2(t)}$, is obtained
from $\ket{\psi_1(t)}$ in Eq.~\eqref{eqn-psi1}
by making the following substitutions:
\begin{eqnarray}
 \ket{1,0}&\to&e^{-i\phi_1}\ket{1,0},\nonumber\\
 \ket{0,1}&\to&e^{-i\phi_2}\ket{0,1},\nonumber
\end{eqnarray}
where
\begin{equation}
 \phi_i=\phi_i(t)\equiv\int_0^t{}dt'\delta\omega_{y,i}(t')\quad(i=1,2)\nonumber
\end{equation}
is the accumulated phase resulting from the stochastic process.
The explicit form of $\ket{\psi_2(t)}$ is:
\begin{eqnarray}
 \ket{\psi_2(t)}
=\frac{1}{2}[
e^{-i\kappa{}t/2}(e^{-i\phi_1}\ket{1,0}+e^{-i\phi_2}\ket{0,1})\nonumber\\
+e^{i\kappa{}t/2}(e^{-i\phi_1}\ket{1,0}-e^{-i\phi_2}\ket{0,1})
].
\label{eqn-psi2}
\end{eqnarray}
The density operator corresponding to this state is:
 $\hat\rho_2(t)\equiv\ket{\psi_2(t)}\bra{\psi_2(t)}$.
Then, the coherence between $\ket{\mathrm{COM}(t)}$ and $\ket{\mathrm{Rock}(t)}$, 
and hence the phonon-hopping coherence,
can be quantified with the following density-matrix element:
 $\bra{\mathrm{COM}(t)}
 \hat\rho_2(t)
 \ket{\mathrm{Rock}(t)}$.
By substituting Eqs.~\eqref{eqn-COMt}, \eqref{eqn-Rockt} and \eqref{eqn-psi2}
into this, the explicit form of the density-matrix element is obtained as
follows:
\begin{eqnarray}
 &&\bra{\mathrm{COM}(t)}
 \hat\rho_2(t)
 \ket{\mathrm{Rock}(t)}\nonumber\\
&=&\frac{1}{8}e^{i\kappa{}t}
[
e^{-i\kappa{}t/2}(e^{-i\phi_1}+e^{-i\phi_2})
+e^{i\kappa{}t/2}(e^{-i\phi_1}-e^{-i\phi_2})
]\nonumber\\
&&\times[
e^{i\kappa{}t/2}(e^{i\phi_1}-e^{i\phi_2})
+e^{-i\kappa{}t/2}(e^{i\phi_1}+e^{i\phi_2})
].
\label{eqn-COMtrho2Rockt}
\end{eqnarray}
We define the common-mode (CM) and differential-mode (DM) frequency shifts,
and the resulting accumulated phases, as follows:
\begin{eqnarray}
 \bar\omega&=&\bar\omega(t)\equiv\frac{1}{2}[\delta\omega_{y,1}(t)+\delta\omega_{y,2}(t)],\nonumber\\
 \Delta\omega&=&\Delta\omega(t)\equiv\frac{1}{2}[\delta\omega_{y,1}(t)-\delta\omega_{y,2}(t)],\nonumber\\
 \bar\phi&=&\bar\phi(t)\equiv\frac{1}{2}[\phi_1(t)+\phi_2(t)]=\int_0^t{}dt'\bar{\omega}(t),\nonumber\\
 \Delta\phi&=&\Delta\phi(t)\equiv\frac{1}{2}[\phi_1(t)-\phi_2(t)]=\int_0^t{}dt'\Delta{\omega}(t).\nonumber
\end{eqnarray}
The CM and DM components introduced above are equivalent to considering
symmetric or anti-symmetric components, respectively.  The
density-matrix element shown in Eq.~\eqref{eqn-COMtrho2Rockt} can then be
represented as follows:
\begin{eqnarray}
 &&\bra{\mathrm{COM}(t)}
 \hat\rho_2(t)
 \ket{\mathrm{Rock}(t)}\nonumber\\
&=&\frac{1}{8}e^{i\kappa{}t}
\left(
-e^{i\Delta\phi}2i\sin\frac{\kappa{}t}{2}
+e^{-i\Delta\phi}2\cos\frac{\kappa{}t}{2}
\right)\nonumber\\
&&\times\left(
e^{i\Delta\phi}2\cos\frac{\kappa{}t}{2}
-e^{-i\Delta\phi}2i\sin\frac{\kappa{}t}{2}
\right).
\label{eqn-COM-rho2-Rock}
\end{eqnarray}
We note that this quantity is dependent on the anti-symmetric or DM
component of phase fluctuations ($\Delta\phi$), but not on the symmetric
or CM component ($\bar\phi$).  This characteristic is relevant to the
selection of noise spectral components, which we investigate later.

Thus, we can connect the phonon-hopping coherence to the time dependence of the phase factors
$e^{\pm{}i\Delta\phi}$, and hence to the difference in the stochastic variables 
$\delta\omega_{y,i}(t)$ ($i=1,2$).

\subsubsection{Effect of dephasing due to electric-potential noise on phonon-hopping coherence}
\label{sec:dephasing-3}

Next, we consider the effect of the stochastic process
in the phase factor $e^{\pm{}i\Delta\phi}$, which leads to dephasing:
\begin{equation}
 e^{\pm{}i\Delta\phi}=\exp\left(\pm{}i\int_0^t{}dt'\Delta\omega(t')\right).\nonumber
\end{equation}
We assume that $\Delta\omega$ behaves as an unbiased stationary Gaussian process  whose correlation time is much shorter 
than the intrinsic coherence time of the local-phonon system
(i.e., that expected in the absence of this stochastic process).
See Appendix~\ref{app:1} for the implications and validity of the assumption.
Although idealized, this assumption provides a reasonable starting point for the analysis.

We define a phase quantity $\varphi$ as the integral of a stochastic variable $f(t)$:
\begin{equation}
 \varphi=\varphi(t)\equiv\int_0^tdt'f(t').\nonumber
\end{equation} 
In Appendix~\ref{app:1} we show that under a similar assumption as
above, the phase factor $e^{i\varphi}$ can
be approximated with an exponential-decay function with the dephasing
rate $\Gamma_\mathrm{d}$ as:
\begin{equation}
 \left\langle{}e^{i\varphi}\right\rangle\simeq{}e^{-\Gamma_\mathrm{d}t},
 \label{eqn-e-i-varphi}
\end{equation}
where
\begin{eqnarray}
 \Gamma_\mathrm{d}&\equiv&\frac{1}{2}S(0),\nonumber\\
 S(\omega)&\equiv&\int_{-\infty}^\infty{}d\tau{}\,e^{i\omega{}\tau}\langle{}f(t)f(t+\tau)\rangle.\nonumber
\end{eqnarray}
Here, $\langle\cdots\rangle$ represents an ensemble average with respect 
to the stochastic process.
$S(\omega)$ is the power spectrum of the stochastic variable $f(t)$
(``noise spectrum''). $S(0)$ formally represents
the zero-frequency component of the power spectrum and effectively represents
its low-frequency component.

By making the following substitutions in Eq.~\eqref{eqn-e-i-varphi}:
\begin{eqnarray}
 f(t)&\to&\pm\Delta\omega(t),\nonumber\\
 \varphi&\to&\pm\Delta\phi\nonumber
\end{eqnarray}
(with the same order of signs),
the phase factors $e^{\pm{}i\Delta\phi}$ can be approximated as:
\begin{equation}
 \left\langle{}e^{\pm{}i\Delta\phi}\right\rangle\simeq{}e^{-\Gamma_{\Delta\omega}t},
  \label{eqn-e-Gamma-Delta-omega}
\end{equation}
where
\begin{eqnarray}
 \Gamma_{\Delta\omega}&\equiv&\frac{1}{2}S_{\Delta\omega}(0),\nonumber\\
 S_{\Delta\omega}(\omega)&\equiv&\int_{-\infty}^\infty{}d\tau{}\,e^{i\omega{}\tau}\langle{}\Delta\omega(t)\Delta\omega(t+\tau)\rangle.
\label{eqn-S-Delta-omega}
\end{eqnarray}
Thus,
the phase factors are approximated using the noise spectrum of the DM frequency shift.
Note that the exponential function
$e^{-\Gamma_{\Delta\omega}t}$ on the right-hand side of Eq.~\eqref{eqn-e-Gamma-Delta-omega} 
remains a real quantity, 
even though it is defined using a complex expression on the left-hand side. 
This is due to the unbiased (i.e., zero-mean) nature of $\Delta\omega$ and $\Delta\phi$.

Using the approximated phase factor given in
Eq.~\eqref{eqn-e-Gamma-Delta-omega}, the ensemble average of the
density-matrix element in Eq.~\eqref{eqn-COM-rho2-Rock} can be greatly
simplified:

\begin{eqnarray}
 &&\bra{\mathrm{COM}(t)}
 \langle\hat\rho_2(t)\rangle
 \ket{\mathrm{Rock}(t)}\nonumber\nonumber\\
&=&\frac{1}{8}e^{i\kappa{}t}
\left(
-\langle{}e^{i\Delta\phi}\rangle{}2i\sin\frac{\kappa{}t}{2}
+\langle{}e^{-i\Delta\phi}\rangle{}2\cos\frac{\kappa{}t}{2}
\right)\nonumber\nonumber\\
&&\times\left(
\langle{}e^{i\Delta\phi}\rangle{}2\cos\frac{\kappa{}t}{2}
-\langle{}e^{-i\Delta\phi}\rangle{}2i\sin\frac{\kappa{}t}{2}
\right)\nonumber\\
&\simeq&\frac{1}{8}e^{i\kappa{}t}
e^{-2\Gamma_{\Delta\omega}t}
\left(
2\cos\frac{\kappa{}t}{2}
-2i\sin\frac{\kappa{}t}{2}
\right)^2\nonumber\\
&=&\frac{1}{2}e^{-2\Gamma_{\Delta\omega}t}.
\label{eqn-2Gamma-Delta-omega}
\end{eqnarray}
Thus, the time-dependence of the density-matrix element
$\bra{\mathrm{COM}(t)} \hat\rho_2(t) \ket{\mathrm{Rock}(t)}$, and hence
the phonon-hopping coherence, is subject to dephasing due to
electric-potential noise, and the decay rate $2\Gamma_{\Delta\omega}$ is
related to the power spectrum of DM frequency shifts,
$S_{\Delta\omega}(\omega)$.

In Appendix~\ref{app:trap-frequencies-second-order-derivative}, we show
that the DM frequency shift $\Delta\omega$ is proportional to the DM
variation for the second-order derivative of the fluctuating electric
potential (denoted as $\Delta\Phi_\mathrm{DM}''$ in
Appendix~\ref{app:trap-frequencies-second-order-derivative}).
Therefore, we can identify $\Delta\omega$ with
$\Delta\Phi_\mathrm{DM}''$ up to a proportionality factor.  Based on
this observation, we denote the noise spectral component
$S_{\Delta\omega}(0)$ in the expression of $\Gamma_{\Delta\omega}$ as
$S_{2,\mathrm{DM}}(0)$.  Here, the subscript ``2'' indicates the
second-order derivative of the electric potential.  Thus,
$\Gamma_{\Delta\omega}$ is represented as follows:
\begin{equation}
 \label{eqn-Gamma-Delta-omega}
 \Gamma_{\Delta\omega}=\frac{1}{2}S_{2,\mathrm{DM}}(0)
\end{equation}

\subsection{Evaluation of phonon-hopping decay rate}
In the previous subsection, we confirmed that phonon-hopping decoherence
can be related to a noise spectral component $S_{2,\mathrm{DM}}(0)$ [or
$S_{\Delta\omega}(0)$].  Next, we explore whether
this quantity can be evaluated based on experimentally observable quantities.

The noise spectral component 
$S_{2,\mathrm{DM}}(0)$ [or $S_{\Delta\omega}(0)$] is associated with 
the low-frequency component
of
the anti-symmetric (DM) fluctuations of the electric potential.
More specifically, it is related
to the second-order derivative of the electric potential,
which corresponds to
the potential curvature or electric-field gradient.
We can consider different types of 
fluctuations and 
quantitatively evaluate their effect experimentally.
Relevant types include fluctuations with different spatial symmetries,
different frequencies,
or
different orders of the electric potential,
i.e.,
first order (electric field)
or second order (potential curvature or electric-field gradient).
We consider here the low-frequency component of  the symmetric (CM) fluctuations
of the electric potential (its second-order derivative),
as well as
components of
symmetric (CM) and anti-symmetric (DM) fluctuations
of the electric potential (its first-order derivative)
at the collective-mode oscillation frequencies.
We show that the former can be experimentally evaluated through 
motional Ramsey interferometry.
Regarding the latter, it is well known that 
heating rates for vibrational modes in ion strings
depend on electric-field noise spectral components at the collective-mode
frequencies.
Conversely, we can evaluate  
electric-field noise spectral components at those frequencies through heating rate measurements.

We also try to find a relation that connects the experimentally
observable quantities.  We make certain simplifying assumptions and
derive a proportionality relation connecting noise spectral components
of different types.

In Sec.~\ref{sec:motional-coherence}, we first focus on
$S_{\bar\omega}(0)$, the low-frequency component of the noise spectrum
of the CM frequency shift.
We show that $S_{\bar\omega}(0)$ is related to motional coherence, which
is typically defined as the coherence between the motional ground state
and the first excited state, i.e., the state with one phonon (either in
local or collective modes).  This motional coherence can be
experimentally determined by performing motional Ramsey interferometry
\cite{roos2008nonlinear}.

Then, in Sec.~\ref{sec:heating} we discuss using heating-rate measurements to determine the noise spectra of the
first-order derivative of the electric potential (i.e., electric
fields), or more correctly, their components at frequencies close to
motional frequencies of the ion string.  We confirm that the heating rates of the COM
and rocking modes are related to CM and DM fluctuations in the
first-order derivative of the electric potential.

Finally, in Sec.~\ref{sec:proportionality} we connect
components of noise spectra with different spatial symmetries and
frequencies based on a simplifying assumption.  We assume a
proportionality relation between the components of noise spectra, from
which $S_{2,\mathrm{DM}}(0)$ and $\Gamma_{\Delta\omega}$ can be
evaluated.

\subsubsection{Effect of dephasing due to electric-potential noise on motional coherence}
\label{sec:motional-coherence} 

Here, we show that motional coherence evaluated in motional Ramsey
interferometry \cite{roos2008nonlinear} is related to the noise spectrum
of the CM frequency shift.

Motional coherence for a single ion can be defined as the coherence
between the ground state $\ket{0}$ and the first excited state $\ket{1}$
of the motional Fock states.  Motional coherence for an ion string with
multiple ions can be defined as that for each collective mode.  Motional
coherence for local vibrational modes may also be defined, though for
simplicity we restrict our treatment here to collective
modes.

In the motional Ramsey interferometry experiment, we first prepare
a superposition of the motional ground state $\ket{0,0}$
and either of $\{\ket{\mathrm{COM}(0)},\ket{\mathrm{Rock}(0)}\}$
using a set of optical pulses (``preparation pulses'').
The time dependence of the generated states without considering the influence of the noise is:
\begin{eqnarray}
 \ket{\psi_3(t)}&=&\frac{1}{\sqrt{2}}(\ket{0,0}+\ket{\mathrm{COM}(t)})\nonumber\nonumber\\
 &=&\frac{1}{\sqrt{2}}\left[\ket{0,0}+e^{i\kappa{}t/2}\frac{1}{\sqrt{2}}(\ket{1,0}+\ket{0,1})\right]\nonumber\\
 \ket{\psi_5(t)}&=&\frac{1}{\sqrt{2}}(\ket{0,0}+\ket{\mathrm{Rock}(t)})\nonumber\\
 &=&\frac{1}{\sqrt{2}}\left[\ket{0,0}+e^{-i\kappa{}t/2}\frac{1}{\sqrt{2}}(\ket{1,0}-\ket{0,1})\right].\nonumber
\end{eqnarray}
Then, the time dependence of the generated states
under the influence of the noise is:
\begin{eqnarray}
 &&\ket{\psi_4(t)}\nonumber\\
&=&\frac{1}{\sqrt{2}}\left[\ket{0,0}+e^{i\kappa{}t/2}\frac{1}{\sqrt{2}}(e^{-i\phi_1}\ket{1,0}+e^{-i\phi_2}\ket{0,1})\right]\nonumber\\
 &&\ket{\psi_6(t)}\nonumber\\
&=&\frac{1}{\sqrt{2}}\left[\ket{0,0}+e^{-i\kappa{}t/2}\frac{1}{\sqrt{2}}(e^{-i\phi_1}\ket{1,0}-e^{-i\phi_2}\ket{0,1})\right].\nonumber\\
\,
\label{eqn-psi4psi6}
\end{eqnarray}
The density operators corresponding to these states are:
 $\hat\rho_4(t)\equiv\ket{\psi_4(t)}\bra{\psi_4(t)}$,
 $\hat\rho_6(t)\equiv\ket{\psi_6(t)}\bra{\psi_6(t)}$.
After the preparation of the superposition state and a time delay of variable length,
another set of optical pulses (``analysis pulses'') is applied, 
which recombines 
the two wave packets into which the probability amplitude was split,
allowing the interference of the two wave packets to be observed.
At this point, the following density-matrix elements can be evaluated through the measurement of
the population:
 $\bra{0,0}\hat\rho_4(t)\ket{\mathrm{COM}(t)}$,
 $\bra{0,0}\hat\rho_6(t)\ket{\mathrm{Rock}(t)}$.
By substituting Eq.~\eqref{eqn-psi4psi6} into these, we obtain:
\begin{eqnarray}
 \bra{0,0}\hat\rho_4(t)\ket{\mathrm{COM}(t)}&=&\frac{1}{4}(e^{-i\phi_1}+e^{-i\phi_2})
  \label{eqn-00-rho4-COM}\\
 \bra{0,0}\hat\rho_6(t)\ket{\mathrm{Rock}(t)}&=&\frac{1}{4}(e^{-i\phi_1}+e^{-i\phi_2}).
  \label{eqn-00-rho4-Rock}
\end{eqnarray}
Therefore, we need to evaluate the phase factors at each ion site,
$e^{-i\phi_1}$ ($i=1,2$), under the influence of the noise.  In
Appendix~\ref{app:3}, we show that,
under the assumption that the magnitude of CM noise is much larger than
that of DM noise,
the power spectrum of $\delta\omega_{y,i}$, $S_{\delta\omega_{y,i}}(\omega)$, 
is approximately equal to $S_{\bar\omega}(\omega)$:
\begin{equation}
 S_{\delta\omega_{y,i}}(\omega)
  \simeq
 S_{\bar\omega}(\omega)
\label{eqn-S-delta-omega-y-i}
\end{equation}
($i=1,2$), where
\begin{equation}
 S_{\delta\omega_{y,i}}(\omega)\equiv\int_{-\infty}^\infty{}d\tau\,{}e^{i\omega{}\tau}\langle{}\delta\omega_{y,i}(t)\delta\omega_{y,i}(t+\tau)\rangle\nonumber
\end{equation}
($i=1,2$) and
\begin{equation}
 S_{\bar\omega}(\omega)\equiv\int_{-\infty}^\infty{}d\tau\,{}e^{i\omega{}\tau}\langle{}\bar\omega(t)\bar\omega(t+\tau)\rangle.
\label{eqn-S-bar-omega}
\end{equation}
Then, the following approximation is valid:
\begin{eqnarray}
 \langle{}e^{-i\phi_i}\rangle\simeq{}e^{-\Gamma_{\bar\omega}t}\quad(i=1,2),
 \label{eqn-e-i-phi-i}
\end{eqnarray}
where
\begin{eqnarray}
 \Gamma_{\bar\omega}&\equiv&\frac{1}{2}S_{\bar\omega}(0).\nonumber
\end{eqnarray}
From Eqs.~\eqref{eqn-00-rho4-COM}, \eqref{eqn-00-rho4-Rock} and \eqref{eqn-e-i-phi-i}: 
\begin{eqnarray}
 \bra{0,0}\langle\hat\rho_4(t)\rangle\ket{\mathrm{COM}(t)}
&\simeq&\bra{0,0}\langle\hat\rho_6(t)\rangle\ket{\mathrm{Rock}(t)}\nonumber\\
&\simeq& \frac{1}{2}e^{-\Gamma_{\bar\omega}t}.
\end{eqnarray}
Thus, motional coherence evaluated in motional Ramsey interferometry is related to
the noise spectrum of the CM frequency shift.

In Appendix~\ref{app:trap-frequencies-second-order-derivative}, we show
that the CM frequency shift $\Delta\omega$ is proportional to the CM
variation for the second-order derivative of the fluctuating electric
potential (denoted as $\Delta\Phi_\mathrm{CM}''$ in
Appendix~\ref{app:trap-frequencies-second-order-derivative}).
Therefore, we can identify $\bar\omega$ with $\Delta\Phi_\mathrm{CM}''$
up to a proportionality factor.  Based on this observation, we denote
the noise spectral component $S_{\bar\omega}(0)$ in the expression of
$\Gamma_{\bar\omega}$ as $S_{2,\mathrm{CM}}(0)$.  Here, the
subscript ``2'' indicates the second-order derivative of the electric
potential.  Thus, $\Gamma_{\bar\omega}$ is represented as
follows:
\begin{equation}
 \label{eqn-Gamma-bar-omega}
 \Gamma_{\bar\omega}=\frac{1}{2}S_{2,\mathrm{CM}}(0)
\end{equation}

\subsubsection{Effect of heating due to electric-potential noise on motional-state populations}
\label{sec:heating}
It is widely known that
electric-field noise
affects motional-state populations by inducing heating of vibrational modes.
The relation between electric-field noise and heating rate is described
in \cite{turchette2000heating}. Here, we review the relevant formalism.
The Hamiltonian for a single ion of charge $q$ and mass $m$ in a
harmonic potential subject to a fluctuating electric field $\epsilon(t)$
is described as:
\begin{equation}
 \hat{H}_1(t)=\hat{H}_{1,0}-q\epsilon(t)\hat{y},
\label{eqn-H-1}
\end{equation}
where $\hat{H}_{1,0}=\hat{p}^2/2m+m\omega_m^2\hat{y}^2/2$ is the harmonic-oscillator
Hamiltonian with trap frequency $\omega_m$.  The heating rate for
the transition from the motional ground state $\ket{0}$ to the first excited
state $\ket{1}$ is obtained to be \footnote{The coefficient for the
heating rate is different from that of Turchette {\it et al.}
\cite{turchette2000heating} by a factor of 2. This is due to the
difference in the definition of the power spectrum by the same
factor. }:
\begin{equation}
 \Gamma_{0\to{1}}=\frac{q^2}{2m\hbar\omega_m}S_E(\omega_m),
\label{eqn-Gamma-0to1}
\end{equation} 
where
\begin{equation}
  S_{E}(\omega)
   \equiv\int_{-\infty}^\infty{}d\tau\,{}
   e^{i\omega{}\tau}\langle{}\epsilon(t)\epsilon(t+\tau)\rangle\nonumber
\end{equation}
is the power spectrum of $\epsilon(t)$.

We extend this formalism to the case of radial motion of a two-ion
string.  The derivation is given in Appendix~\ref{app:4} and the results are
reviewed here.  We define the CM and DM variation of the electric field
as follows:
\begin{eqnarray}
 \epsilon_\mathrm{CM}(t)&\equiv&\frac{1}{\sqrt{2}}(\epsilon_1(t)+\epsilon_2(t)),\nonumber\\
 \epsilon_\mathrm{DM}(t)&\equiv&\frac{1}{\sqrt{2}}(\epsilon_1(t)-\epsilon_2(t)),\nonumber
\end{eqnarray}
where $\epsilon_i(t)$ ($i=1,2$) is the fluctuating electric field at the
$i$th ion site.  Then, the heating rates for the radial COM and Rocking
modes are obtained to be:
\begin{eqnarray}
 \Gamma_{\mathrm{COM},0\to{}1}
  &=&\frac{q^2}{2m\hbar\omega_\mathrm{COM}}S_{1,\mathrm{CM}}(\omega_\mathrm{COM}),
  \label{eqn-Gamma-COM-0-1}\\
 \Gamma_{\mathrm{Rock},0\to{}1}
  &=&\frac{q^2}{2m\hbar\omega_\mathrm{Rock}}S_{1,\mathrm{DM}}(\omega_\mathrm{Rock}),
  \label{eqn-Gamma-Rock-0-1}
\end{eqnarray}
where
\begin{eqnarray}
  S_{1,\mathrm{CM}}(\omega)
   &\equiv&\int_{-\infty}^\infty{}d\tau\,{}
   e^{i\omega{}\tau}\langle{}\epsilon_\mathrm{CM}(t)\epsilon_\mathrm{CM}(t+\tau)\rangle,\nonumber\\
  S_{1,\mathrm{DM}}(\omega)
   &\equiv&\int_{-\infty}^\infty{}d\tau\,{}
   e^{i\omega{}\tau}\langle{}\epsilon_\mathrm{DM}(t)\epsilon_\mathrm{DM}(t+\tau)\rangle.\nonumber
\end{eqnarray}
Thus, the heating rates for radial motion in a two-ion string are related
to the noise spectra of the electric field (the first derivative of the electric potential).

\subsubsection{Proportionality relation connecting noise spectral components}
\label{sec:proportionality}
We have confirmed that various components of noise spectra
are related to experimentally observable quantities.
Here, we introduce a simplifying assumption 
for the proportionality relation of those noise spectral components,
which can be used to evaluate a component distinct from the others.

In an ion trap, various factors can contribute as sources of
electric-potential noise.  For example, RF or DC electrodes and connected circuitry
may generate noise, and materials attached to the
electrodes may generate electric-field noise, leading to anomalous
heating \cite{Deslauriers2006}. In addition, other external sources may also
contribute.  Here we consider a simplified situation where the noise
that acts on the trapped ions is dominated by a single noise source
with a particular spectrum, which generates a fluctuating electric
potential with a particular spatial variation.  We suppose that the
noise source has nearly ideal properties as noise (i.e., an unbiased
stationary Gaussian process with a short correlation time, as stated
above).  Under such conditions, we assume the following proportionality
relation of the noise spectral components:
\begin{eqnarray}
 &&S_{2,\mathrm{CM}}(0):
 S_{2,\mathrm{DM}}(0)\nonumber\\
 &=&
 S_{1,\mathrm{CM}}(\omega_\mathrm{COM}):
 S_{1,\mathrm{DM}}(\omega_\mathrm{Rock}).
 \label{eqn-proportionality1}
\end{eqnarray}

The justification for the relation in Eq.~\eqref{eqn-proportionality1} is
provided in Appendix~\ref{app:5}.  It is based on the following
observations: (i) the spatial profile of the electric potential is
similar for different frequencies and (ii) the local
potentials around the two ions are similar, due to the fact
that the electric potential varies smoothly over the distance between
the ions.  In the following, we use the relation as given above.

Based on the proportionality relation in Eq.~\eqref{eqn-proportionality1}
for the noise spectra components, we can derive a relation for the decay
rates.  Using the relations
$\Gamma_{\bar\omega}=S_{2,\mathrm{CM}}(0)/2$ [Eq.~\eqref{eqn-Gamma-bar-omega}],
$\Gamma_{\Delta\omega}=S_{2,\mathrm{DM}}(0)/2$ [Eq.~\eqref{eqn-Gamma-Delta-omega}],
$\Gamma_{\mathrm{COM},0\to{}1}
=({q^2}/{2m\hbar\omega_\mathrm{COM}})S_{1,\mathrm{CM}}(\omega_\mathrm{COM})$ [Eq.~\eqref{eqn-Gamma-COM-0-1}],
and $\Gamma_{\mathrm{Rock},0\to{}1}
=({q^2}/{2m\hbar\omega_\mathrm{Rock}})S_{1,\mathrm{DM}}(\omega_\mathrm{Rock})$ [Eq.~\eqref{eqn-Gamma-Rock-0-1}],
it is obtained as:
\begin{equation}
 \Gamma_{\bar\omega}:
  \Gamma_{\Delta\omega}
  \approx
  \Gamma_{\mathrm{COM},0\to{}1}:
  \Gamma_{\mathrm{Rock},0\to{}1},
\end{equation} 
where $\omega_\mathrm{COM}/\omega_\mathrm{Rock}\approx1$ is used.
From this, the phonon-hopping decay rate, 
$2\Gamma_{\Delta\omega}$, is represented as follows:
\begin{equation}
 \label{eqn-2-Gamma-Delta-omega}
  2\Gamma_{\Delta\omega}
  \approx
  \frac{\Gamma_{\mathrm{Rock},0\to{}1}}
  {\Gamma_{\mathrm{COM},0\to{}1}}
 2\Gamma_{\bar\omega}
\end{equation} 
Thus, the phonon-hopping decay rate
is represented using other experimentally observable quantities.

\subsection{Evaluation of phonon-hopping decay rate due to electric-potential noise}
\label{sec:eval-decay-rate-noise}
To evaluate the phonon-hopping decay rate due to electrical-potential
noise in Eq.~\eqref{eqn-2-Gamma-Delta-omega}, we independently measured
all the quantities on the right-hand side of
Eq.~\eqref{eqn-2-Gamma-Delta-omega} under identical trap conditions.
The decay rate for motional coherence, $\Gamma_{\bar\omega}$, was
determined from independent Ramsey-type measurements performed on the
COM and rocking motional modes under the same experimental conditions as
the phonon-hopping experiments. 
A coherent superposition of the motional ground state and the
first-excited state of either the COM or the rocking mode was prepared
using a $\pi/2$ pulse applied on the motional sideband. After a variable
free-evolution time, a second $\pi/2$ pulse was applied to map the
accumulated phase onto the internal-state population.

From exponential fits to the decay of the Ramsey fringe contrast, the
dephasing time $T_{\mathrm{Ramsey},m}$ ($m =
\mathrm{COM},\mathrm{Rock}$) was obtained, yielding
$\Gamma_{\mathrm{Ramsey},m} = 1/T_{\mathrm{Ramsey},m}$.  For the COM
mode, we performed two sets of measurements, obtaining coherence
times of $\{0.84\pm0.33, 1.10\pm0.33\}$ ms.  The quoted uncertainties
reflect the statistical errors of the exponential fits.  For the rocking
mode, we performed two sets of measurements, obtaining coherence
times of $\{1.57\pm0.44, 1.53\pm0.46\}$ ms.  The results for the
rocking mode show longer decay times, though it is not clear at this
point whether this difference is statistically significant, and its
origin remains unknown.  From the results of our simple analysis
[Eqs.~\eqref{eqn-00-rho4-COM}, \eqref{eqn-00-rho4-Rock}, whose
expressions in the right-hand sides are identical], motional Ramsey
experiments with COM and rocking modes are expected to give the same
decay time, acknowledging a possible slight difference between the two types of
measurements that cannot be explained by this analysis.  As a first
approximation, we assume that the observed difference arises from
statistical fluctuations rather than a systematic effect, and estimate
the decay rate by simply averaging the values for the two types of
measurements.  The obtained value is $1.25\pm0.20$ ms, and we use this as the value of $\Gamma_{\bar\omega}^{-1}$.

The heating rates $\Gamma_{\mathrm{COM},0\to1}$ and
$\Gamma_{\mathrm{Rock},0\to1}$ were obtained from standard
motional-heating measurements
\cite{turchette2000heating,savard1997laser}. After sideband cooling to
near the motional ground state, a variable delay time was inserted and
the mean phonon numbers of the COM and rocking modes were measured using
red--blue sideband thermometry. The mean phonon numbers increase linearly
as $\bar n_m(t) \simeq \Gamma_{m,0\to1} t$ ($m =
\mathrm{COM},\mathrm{Rock}$), from which the heating rates were
extracted by linear fits to the measured data.

For the COM mode, a heating rate of $\Gamma_{\mathrm{COM},0\to1} = (8.0
\pm 2.1)\times10^{-2}~\mathrm{quanta/ms}$ was measured, corresponding to
a heating time of $12.5_{-2.6}^{+4.3}~\mathrm{ms}$. Here, the quoted
uncertainties represent the statistical errors of the linear fits. For
the rocking mode, we obtain a heating rate of
$\Gamma_{\mathrm{Rock},0\to1} =
(1.21\pm0.79)\times10^{-3}\,\mathrm{quanta/ms}$, corresponding to a
heating time of $830_{-330}^{+1550}~\mathrm{ms}$.

Using these independently measured quantities and substituting them into
the right-hand side of Eq.~\eqref{eqn-2-Gamma-Delta-omega}, we evaluated
$2\Gamma_{\Delta\omega}=0.022^{+0.023}_{-0.013}$ ms$^{-1}$.

\section{Results of numerical simulations}
In this section, we discuss the results of numerical simulations based
on the contents in the previous two sections and compare them with the
experimental results.

As discussed in
Sec.~\ref{sec:principles-inter-mode-couplings}, the coherent
phonon-hopping dynamics is subject to decoherence arising
from inter-mode coupling due to Kerr nonlinearity.  
The phonon-hopping time profile that takes this effect into account 
is given in Eq.~\eqref{eqn-hKerr} as $h_\mathrm{Kerr}$(t).
To account for this and the effect of electric-potential
noise at the same time, the following time profile can be used:
\begin{equation} \label{eq:h_total}
h_{\mathrm{total}}(t)
=
h_{\mathrm{Kerr}}(t)\,
\exp(-t/T_\mathrm{noise}),
\end{equation}
where $T_\mathrm{noise}\equiv(2\Gamma_{\Delta\omega})^{-1}$.  Based on
the two time profiles given in Eqs.~\eqref{eqn-hKerr} and
\eqref{eq:h_total}, we performed two types of numerical simulations: (i)
a Kerr-only model incorporating the nonlinear coupling between the
rocking and stretch modes [Eq.~\eqref{eqn-hKerr}] and (ii) the 
combined Kerr and electric-potential noise model (Kerr+noise)  of Eq.~(\ref{eq:h_total}), which includes both the
Kerr-induced inter-mode coupling and the noise-induced dephasing.  For
each model we used sets of values for the inter-ion distance $d_0$ and
the radial trap frequency $\omega_y$ that closely match the experimental
conditions.  Figs.~\ref{fig:decay-time} and
\ref{fig:number-of-oscillations} show the results for the Kerr-only model [case (i)] as
green triangles and the results for the Kerr+noise model [case (ii)] as black
crosses with error bars.
In addition, 
numerically simulated results for only electric-potential
noise are also indicated using blue dashed curves (central value) and
blue dotted curves (lower and upper bounds of the confidence interval).

It can be seen from Figs.~\ref{fig:decay-time} and
\ref{fig:number-of-oscillations} that the numerical results for the Kerr-only model [case (i)],
except for Fig.~\ref{fig:number-of-oscillations}(b),
show similar global trends to the experimental results
with respect to
the dependence on the two trap parameters (inter-ion distance and radial trap
frequency). However, a quantitative correspondence is lacking
in this case, and the numerically simulated results significantly
overestimate the absolute values of decay time and number of
oscillations.  On the other hand, the numerical results for the Kerr+noise model [case (ii)] 
are closer to the experimental results for
both decay time (Fig.~\ref{fig:decay-time}) and 
number of oscillations (Fig.~\ref{fig:number-of-oscillations}).

The lack of quantitative correspondence for the Kerr-only model [case (i)] indicates that the
experimentally observed phonon-hopping decoherence cannot be explained
by Kerr-type nonlinear mode coupling alone.  Nevertheless, the
qualitative agreement in the global trends with respect to the two trap
parameters, namely the inter-ion distance and the radial trap frequency,
suggests that this mechanism still plays an important role in the
experimental observations.  
The quantitative correspondence between the unified Kerr+noise model [case (ii)]
and the experimental data supports the conjecture that both Kerr-induced
frequency dispersion and electric-field-induced curvature noise
contribute simultaneously to the phonon-hopping decoherence.  

\section{Discussion}

Studies of coherence in quantum systems typically focus on the lifetime
of internal-state superpositions. In contrast, the coherence considered
here is a form of inter-mode coherence, which characterizes the relative
phase stability between different vibrational modes. In the
phonon-hopping process, this coherence determines how faithfully the COM
and rocking modes remain phase-locked while exchanging
excitations. Maintaining such coherence is essential for implementing
multi-mode quantum operations and for exploiting radial phonons as
carriers of quantum information in trapped-ion systems.

It is valuable to discuss possible ways to improve the phonon-hopping
coherence with respect to the two decoherence mechanisms considered in
this study. For inter-mode coupling due to Kerr nonlinearity treated in
Sec.~\ref{sec:inter-mode-couplings}, 
Eq.~\eqref{eqn-hKerr} shows that the phonon-hopping time
profile is determined by the phonon-number distribution of the axial
stretching mode $P(n_s)$, which weights each sinusoidal component in the
sum; the frequency of each component is shifted by $\delta\omega_r=\chi n_s$
[Eq.~\eqref{eqn-delta-omega-r}].  Therefore, it is expected that reducing the residual axial
thermal distribution can directly mitigate the Kerr-induced dispersion.
For our present conditions, due to the relatively weak confinement along
the axial direction ($\omega_z/2\pi\approx50$--$200$ kHz), the axial motion
does not enter the Lamb-Dicke regime even at the Doppler limit,
and only Doppler cooling was performed in that direction.  A practical
strategy to mitigate the Kerr-induced dispersion is to reduce $\langle
n_s\rangle$ by using sub-Doppler cooling mechanisms (e.g.,
electromagnetically induced transparency cooling \cite{Roos2000}) in the
axial direction, so that the motion in that direction approaches the
ground state.
For the electric-potential noise treated in Sec.~\ref{sec:elec-pot-noise},
the relevant decay rate is $2\Gamma_{\Delta \omega}$, which is
associated with the low-frequency component of the DM variation for the
second-order derivative of the fluctuating electric potential.  This may
be reduced by mitigating technical noise in the RF and DC control
circuitry via, e.g., improving filtering and grounding.

In Figs.~\ref{fig:decay-time} and
\ref{fig:number-of-oscillations},
we still find discernible differences between the experimental results and numerically simulated results for the Kerr+noise model.
Generally,
the experimental results show
longer decay times or larger numbers of oscillations than those
for the Kerr+noise model, and this trend is stronger in the case of
the dependence on the radial trap frequency.
Possible reasons for this difference include
imperfect modeling of noise factors,
imperfect estimation of dephasing time and heating rates,
imperfect estimation and variations of the axial thermal distribution,
drifts in the trap potentials, and variations in the noise environment
associated with different DC settings.
In Fig.~\ref{fig:number-of-oscillations}(b), the experimental data show
monotonically increasing behavior, whereas the numerically simulated
results for the Kerr+noise model show an almost flat or slightly
decreasing behavior.  The reason for this difference is
not currently understood, but possible reasons include imperfect
estimation and variations of the axial thermal distribution.

We confirmed in Sec.~\ref{sec:formalism-for-radial-mode} that
phonon-hopping coherence for radial motion in a two-ion string is
equivalent to the coherence between $|{\rm COM}(t)\rangle$ and $|{\rm
Rock}(t)\rangle$.  The latter is mathematically expressed using
$\bra{\mathrm{COM}(t)} \langle\hat\rho_2(t)\rangle
\ket{\mathrm{Rock}(t)}$ in Eq.~\eqref{eqn-2Gamma-Delta-omega}.  We note
that this quantity can be evaluated in an alternative manner using
motional Ramsey interferometry \cite{roos2008nonlinear}, where a
superposition of $\ket{\mathrm{COM}}(0)\rangle$ and
$\ket{\mathrm{Rock}}(0)$ is first generated in the preparation stage, and a
corresponding procedure is used in the analysis stage.  It suffices in
this case to use two mapping sideband $\pi$ pulses resonant with
$\ket{0}\leftrightarrow\ket{\mathrm{COM}}(t)\rangle$ and
$\ket{0}\leftrightarrow\ket{\mathrm{Rock}}(t)\rangle$, along with a
$\pi/2$ pulse resonant with the carrier transition, for each of the
preparation and analysis stages in the motional Ramsey interferometry.
In the present work, we did not perform a dedicated motional Ramsey
measurement along this line, principally because
applying the required preparation and analysis pulses is
more complex than the other measurements in the present work
and
the results expected if we performed such a measurement
would be equivalent to those obtained in the measurement of phonon
hopping.

Finally, we note that the decoherence factors and evaluation methods
relevant to this study are not limited to radial local phonons in a
linear ion string. Related studies on local phonons in independently
controlled traps (e.g., multi-well potentials or trap arrays) have
demonstrated coherent energy exchange among separated ions in
double-well potentials \cite{Brown2011,Harlander2011}, tunable
interactions and entanglement of ions in separate potential wells
\cite{Wilson2014}, coherent motional coupling enabled by controlled
electric potentials \cite{Hou2024}, and two-dimensional networks of
vibrational modes in microtrap arrays
\cite{Hakelberg2019,Kiefer2019}. In these platforms, local-phonon
coherence can likewise be affected by Coulomb-induced couplings and
electric-potential noise.  Similar studies can be performed with these
platforms and may help to improve coherence therein.

\section{Conclusion}

We have presented a systematic experimental investigation
of phonon-hopping coherence in a two-ion string and performed relevant
theoretical and numerical analyses. In the experiment, by
varying the inter-ion distance and radial trap frequency, we
identified clear dependencies of the phonon-hopping decay time and the
number of oscillations on global trap parameters. While a
Kerr-type nonlinear mode coupling accounts for the qualitative trends,
it does not quantitatively reproduce the measured coherence.
Incorporating the effect of electric-potential noise substantially
reduced this discrepancy: a combined Kerr+noise model that includes both Kerr
dynamics and noise-induced dephasing reproduces the experimental data
well across all parameter regimes.  
These results indicate the importance
of controlling decoherence factors such as nonlinear mode couplings and electric-potential noise
for preserving coherence in phonon-mediated quantum processes. 
The identification and quantitative estimation of multiple decoherence
channels may enable more robust multi-mode quantum simulations and
scaling of phonon-based quantum information processing in trapped-ion
systems.

\appendix
\section{Derivation of time dependence of stochastic phase factor}
\label{app:1}

We review here the finding that the ensemble average of the phase factor
$e^{i\varphi}$ with a phase $\varphi$, which has a stochastic origin,
reduces to an exponential decay under certain conditions
\cite{Bergli2009decoherence,Degen2017,Preskill2006}.

We define a phase quantity $\varphi$ as the integral of a stochastic
variable $f(t)$:
\begin{equation}
 \label{eqn-varphi-2}
 \varphi=\varphi(t)\equiv\int_0^tdt'f(t').
\end{equation} 
$f(t)$ is assumed to be a stationary Gaussian process, that is, the
statistical properties of $f(t)$ are determined by the ensemble average
and the two-time correlation function, where the former does not depend
on time and the latter depends only on the time difference.
Stationarity allows us to consider the spectrum of $f(t)$ as the Fourier
transform of the two-time correlation function.  We also assume that
$f(t)$ is unbiased, i.e., the ensemble average is equal to zero.
Therefore, the statistical properties are determined only by the
two-time correlation function (or the spectrum).  Furthermore, we assume
that the correlation time for $f(t)$ is sufficiently short compared with
the intrinsic coherence time for the system (i.e., the coherence time in
the absence of the stochastic process under consideration).

Since $\varphi(t)$ is the integral of $f(t)$, it
is linearly dependent on the values of $f(t)$ and therefore inherits the Gaussian property of $f(t)$.  In addition,
$\varphi(t)$ is regarded as unbiased, similar to $f(t)$.  The
statistical properties of $\varphi(t)$ can thus be described (at each $t$) by
the Gaussian probability distribution function:
\begin{equation}
 p(\varphi)=\frac{1}{\sqrt{2\pi\langle\varphi^2\rangle}}
  \exp\left(-\frac{\varphi^2}{2\langle\varphi^2\rangle}\right),
\end{equation}
where $\langle\cdots\rangle$ represents an ensemble average with respect
to the stochastic process, and:
\begin{eqnarray}
 \langle{}e^{i\varphi}\rangle
  &=&\int_{-\infty}^{\infty}d\varphi\,
  p(\varphi)
  e^{i\varphi}\nonumber\\
 &=&
  \frac{1}{\sqrt{2\pi\langle\varphi^2\rangle}}
  \int_{-\infty}^{\infty}d\varphi\,
  e^{-{\varphi^2}/{2\langle\varphi^2\rangle}}
  e^{i\varphi}.
\label{eqn-av-e-i-varphi}
\end{eqnarray}
This is similar to the Fourier transform of a Gaussian function.
For a Gaussian function:
\begin{equation}
 \label{eqn-g-varphi}
  g(\varphi)=
 \frac{1}{\sqrt{\langle\varphi^2\rangle}}
  \exp\left(-\frac{\varphi^2}{2\langle\varphi^2\rangle}\right),
\end{equation}
and its Fourier transform is also a Gaussian function, which is expressed as:
\begin{eqnarray}
 G(\nu)&=&
  \frac{1}{\sqrt{2\pi}}
  \int_{-\infty}^{\infty}d\varphi\,
  g(\varphi)
  e^{-i\nu\varphi}\nonumber\\
 &=&
  e^{-\nu^2\langle\varphi^2\rangle/2}.
  \label{eqn-G-nu}
\end{eqnarray}
By substituting Eq.~\eqref{eqn-g-varphi} into Eq.~\eqref{eqn-G-nu}:
\begin{equation}
  \frac{1}{\sqrt{2\pi\langle\varphi^2\rangle}}
  \int_{-\infty}^{\infty}d\varphi\,
  e^{-{\varphi^2}/{2\langle\varphi^2\rangle}}
  e^{-i\nu\varphi}
  =
  e^{-\nu^2\langle\varphi^2\rangle/2},
\end{equation}
and by setting $\nu$ as $-1$:
\begin{equation}
  \frac{1}{\sqrt{2\pi\langle\varphi^2\rangle}}
  \int_{-\infty}^{\infty}d\varphi\,
  e^{-{\varphi^2}/{2\langle\varphi^2\rangle}}
  e^{i\varphi}
  =
  e^{-\langle\varphi^2\rangle/2}.
  \label{eqn-e-minus-varphi2-2}
\end{equation}
Thus, using Eqs.~\eqref{eqn-av-e-i-varphi} and
\eqref{eqn-e-minus-varphi2-2}:
\begin{equation}
 \langle{}e^{i\varphi}\rangle=
  e^{-\langle\varphi^2\rangle/2}. 
  \label{eqn-av-e-i-varphi-2}
\end{equation}

The relation in Eq.~\eqref{eqn-av-e-i-varphi-2} was obtained using the
properties of the noise distribution as an unbiased Gaussian
distribution.  Next, we also used the dynamical or time-dependent
properties of $f(t)$, i.e., the stationarity and short correlation
time, to determine those of $\langle\varphi^2\rangle$ and
$\langle{}e^{i\varphi}\rangle$.

Using the definition of $\varphi$ in Eq.~\eqref{eqn-varphi-2},
$\langle\varphi^2\rangle$ at time $T$ is expressed as follows:
\begin{eqnarray}
 \langle\varphi^2\rangle
&=&
 \left\langle
 \int_0^Tdt_1\,f(t_1)
 \int_0^Tdt_2\,f(t_2)
 \right\rangle
\nonumber\\&=&
 \int_0^Tdt_1
 \int_0^Tdt_2
 \langle
 f(t_1)f(t_2)
 \rangle
\nonumber\\&=&
 \int_0^Tdt_1
 \int_0^Tdt_2
 \langle
 f(t)f(t+\tau)
 \rangle,
\label{eqn-av-varphi2}
\end{eqnarray}
where we made the following substitutions in the integrand in the last line: $t_1\to{}t$,
$t_2-t_1\to\tau$.

From the stationarity of $f(t)$, we can consider the power spectrum 
of $f(t)$, which is equal to the Fourier transform of the correlation
function:
\begin{equation}
 S(\omega)\equiv
  \int_{-\infty}^\infty{}d\tau\,
  e^{i\omega{}\tau}\langle{}f(t)f(t+\tau)\rangle.
  \label{eqn-S-omega-2}
\end{equation}
Taking the inverse Fourier transform of both sides of
Eq.~\eqref{eqn-S-omega-2}, we have:
\begin{equation}
 \frac{1}{2\pi}
  \int_{-\infty}^\infty{}d\omega\,
  e^{-i\omega\tau}
  S(\omega)
  =
  \langle{}f(t)f(t+\tau)\rangle.
  \label{eqn-av-f-f}
\end{equation}
By substituting Eq.~\eqref{eqn-av-f-f} into Eq.~\eqref{eqn-av-varphi2} and
using $\tau=t_2-t_1$, we obtain:
\begin{eqnarray}
 \langle\varphi^2\rangle
&=&
\frac{1}{2\pi}
\int_0^Tdt_1
\int_0^Tdt_2
\int_{-\infty}^\infty{}d\omega\,
e^{-i\omega(t_2-t_1)}
S(\omega)
\nonumber\\&=&
\frac{1}{2\pi}
\int_{-\infty}^\infty{}d\omega\,
S(\omega)
\int_0^Tdt_1\,
e^{i\omega{}t_1}
\int_0^Tdt_2\,
e^{-i\omega{}t_2}
\nonumber\\&=&
\frac{1}{2\pi}
\int_{-\infty}^\infty{}d\omega\,
S(\omega)
\frac{4}{\omega^2}
\sin^2
\frac{\omega{}T}{2}
\label{eqn-av-varphi2-2}
\end{eqnarray}

The function $({4}/{\omega^2}) \sin^2 ({\omega{}T}/{2})$ can be
interpreted as the filter function associated with free-induction decay
\cite{Degen2017}, which weights the noise spectrum $S(\omega)$ in
Eq.~\eqref{eqn-av-varphi2-2}.  This filter function effectively selects
a frequency window of width $\sim1/T$ around $\omega=0$, where the
filter function has a height proportional to $T^2$.

Here, we consider the typical magnitude of $T$ (the time at which
$\langle\varphi^2\rangle$ and $\langle{}e^{i\varphi}\rangle$ is
evaluated) in comparison with the correlation time of $f(t)$.  In our
study of phonon-hopping coherence and its decay, the typical values of
$T$ are comparable to, or smaller than, the intrinsic coherence time for
the system (the coherence time in the absence of the stochastic process under
consideration), while not differing from it by orders of magnitude.  On
the other hand, we assume that the correlation time of $f(t)$ is
sufficiently short compared with the intrinsic coherence time for the
system.  Therefore, it is reasonable to assume that the correlation time for
$f(t)$ is also sufficiently short compared with $T$ and, conversely, 
$T$ can be assumed to be sufficiently long compared with the correlation time.

If $T$ is taken to be sufficiently long compared with the correlation
time for the stochastic variable $f(t)$, the noise spectrum $S(\omega)$
can be considered to be almost constant over the width of the filter
function $\sim1/T$.
Then, the filter function is effectively considered
to be proportional to a delta function.
This is mathematically confirmed from the following relation:
\begin{equation}
 \lim_{T\to\infty}
\frac{4}{\omega^2}
\sin^2
\frac{\omega{}T}{2}
=
2\pi{}T\delta(\omega).
\label{eqn-lim-FF}
\end{equation}
Therefore, in the limit that $T$ is sufficiently long compared with the
correlation time of $f(t)$, we use Eqs.~\eqref{eqn-av-varphi2-2},
\eqref{eqn-lim-FF} and \eqref{eqn-av-e-i-varphi-2} to derive the
following relations:
\begin{eqnarray}
 \langle\varphi^2\rangle&\simeq&TS(0),\\
 \langle{}e^{i\varphi}\rangle&\simeq&e^{-TS(0)/2}=e^{-\Gamma_\mathrm{d}T},
\end{eqnarray}
where
\begin{equation}
 \Gamma_\mathrm{d}\equiv\frac{1}{2}S(0).
\end{equation}
Thus, the time dependence of the ensemble average of the phase factor
is represented as an exponential-decay function
with the decay rate $\Gamma_\mathrm{d}$.

\section{Relation between radial trap frequencies and second-order derivative of the electric potential}
\label{app:trap-frequencies-second-order-derivative}

In this appendix, we consider the quantitative relation between the trap
frequency shifts and fluctuating second-order derivative of the electric
potential, and confirm that the CM and DM frequency shifts are
proportional to CM and DM fluctuations in the second-order derivative of
the electric potential, respectively.  This relation is implicitly
assumed in Secs.~\ref{sec:dephasing} and \ref{sec:motional-coherence},
and is demonstrated explicitly in this appendix.
The potential energy for the two-ion string (in the classical notation) is described as:
\begin{equation}
 \sum_{i=1}^{2}\Phi_{y,i}(y_i)=\sum_{i=1}^{2}\frac{1}{2}m(\omega_y+\delta\omega_{y,i})^2{y}_i^2,
\end{equation}
where $\Phi_{y,i}(y_i)$ ($i=1,2$) is the electric potential for the $i$th ion along the radial $y$ direction.
We assume $\delta\omega_{y,i}\ll\omega_y$,
and expand the quadratic form in the expression up to the first order of $\delta\omega_{y,i}$:
\begin{eqnarray}
 \sum_{i=1}^{2}\Phi_{y,i}(y_i)
&\simeq&\sum_{i=1}^{2}\frac{1}{2}m(\omega_y^2+2\omega_y\delta\omega_{y,i}){y}_i^2\nonumber\\
&=&\sum_{i=1}^{2}(\Phi_{y,i,0}+\Delta\Phi_{y,i}),
\end{eqnarray}
where
\begin{eqnarray}
 \Phi_{y,i,0}(y_i)&\equiv&\frac{1}{2}m\omega_y^2{y}_i^2,\nonumber\\
 \Delta\Phi_{y,i}(y_i)&\equiv&m\omega_y\delta\omega_{y,i}{y}_i^2\nonumber
\end{eqnarray}
($i=1,2$).

We define the CM and DM variation for the second-order derivative of the electric potential:
\begin{eqnarray}
 \Delta\Phi_\mathrm{CM}''&\equiv&\frac{1}{2}
  \left[
   \frac{d^2}{dy_1^2}\Delta\Phi_{y,1}(y_1)
   +\frac{d^2}{dy_2^2}\Delta\Phi_{y,2}(y_2)  
  \right],\nonumber\\
 \Delta\Phi_\mathrm{DM}''&\equiv&\frac{1}{2}
  \left[
   \frac{d^2}{dy_1^2}\Delta\Phi_{y,1}(y_1)
   -\frac{d^2}{dy_2^2}\Delta\Phi_{y,2}(y_2)
  \right],\nonumber
\end{eqnarray}
which are obtained to be:
\begin{eqnarray}
 \Delta\Phi_\mathrm{CM}''
&=&\frac{1}{2}m\omega_y(\delta\omega_{y,1}+\delta\omega_{y,2})\nonumber\\
&=&m\omega_y\bar\omega,\\
 \Delta\Phi_\mathrm{DM}''
&=&\frac{1}{2}m\omega_y(\delta\omega_{y,1}-\delta\omega_{y,2})\nonumber\\
&=&m\omega_y\Delta\omega.
\end{eqnarray}
From this, we confirm the proportionality relations:
\begin{eqnarray}
 \Delta\Phi_\mathrm{CM}''&\propto&\bar\omega,\\
 \Delta\Phi_\mathrm{DM}''&\propto&\Delta\omega,
\end{eqnarray}
Thus, we can identify $\Delta\Phi_\mathrm{CM}''$ with $\bar\omega$ and
$\Delta\Phi_\mathrm{DM}''$ with $\Delta\omega$, up to the
proportionality factor $m\omega_y$.

\section{Approximate expression for noise spectrum at each ion site in terms of common-mode noise spectrum}
\label{app:3}

Here, we detail Eq.~\eqref{eqn-S-delta-omega-y-i} in the main text ($i=1,2$):
\begin{equation}
 S_{\delta\omega_{y,i}}(\omega)
  \simeq
 S_{\bar\omega}(\omega).\nonumber
\label{eqn-S-delta-omega-y-i-2}
\end{equation}
$S_{\bar\omega}(\omega)$ in Eq.~\eqref{eqn-S-bar-omega} and
$S_{\Delta\omega}(\omega)$ in Eq.~\eqref{eqn-S-Delta-omega}
can be represented as follows:
\begin{eqnarray}
 &&S_{\bar\omega}(\omega)\nonumber
  =\int_{-\infty}^\infty{}d\tau\,
  e^{i\omega{}\tau}
  \frac{1}{4}
  [
  \langle{}\delta\omega_{y,1}(t)\delta\omega_{y,1}(t+\tau)\rangle\nonumber\\
 &&+2\langle{}\delta\omega_{y,1}(t)\delta\omega_{y,2}(t+\tau)\rangle
  +\langle{}\delta\omega_{y,2}(t)\delta\omega_{y,2}(t+\tau)\rangle
  ],
  \nonumber
  \\
  \label{eqn-S-bar-omega-2}
 \\
 &&S_{\Delta\omega}(\omega)\nonumber
  =\int_{-\infty}^\infty{}d\tau\,
  e^{i\omega{}\tau}
  \frac{1}{4}
  [
  \langle{}\delta\omega_{y,1}(t)\delta\omega_{y,1}(t+\tau)\rangle\nonumber\\
 &&-2\langle{}\delta\omega_{y,1}(t)\delta\omega_{y,2}(t+\tau)\rangle
  +\langle{}\delta\omega_{y,2}(t)\delta\omega_{y,2}(t+\tau)\rangle
  ].
  \nonumber
  \\
  \label{eqn-S-Delta-omega-2}
\end{eqnarray}
Here, we used the relation
$\langle{}\delta\omega_{y,1}(t)\delta\omega_{y,2}(t+\tau)\rangle=
\langle{}\delta\omega_{y,1}(t+\tau)\delta\omega_{y,2}(t)\rangle$ due to
the symmetry of the two-time correlation function and its
$t$-invariance.  Using Eqs.~\eqref{eqn-S-bar-omega-2} and
\eqref{eqn-S-Delta-omega-2}:
\begin{eqnarray}
 &&S_{\bar\omega}(\omega)
  +S_{\Delta\omega}(\omega)\nonumber\\
  &=&\int_{-\infty}^\infty{}d\tau\,
  e^{i\omega{}\tau}
  \frac{1}{2}
  [
  \langle{}\delta\omega_{y,1}(t)\delta\omega_{y,1}(t+\tau)\rangle\nonumber\\
  &&+\langle{}\delta\omega_{y,2}(t)\delta\omega_{y,2}(t+\tau)\rangle
  ]
  \nonumber\\
  &=&\frac{1}{2}
  [
  S_{\delta\omega_{y,1}}(\omega)+S_{\delta\omega_{y,2}}(\omega)
  ].
  \label{eqn-appC1}
\end{eqnarray}
We assume that the noise spectra for the two ions
are approximately equal:
\begin{equation}
S_{\delta\omega_{y,1}}(\omega)\simeq{}S_{\delta\omega_{y,2}}(\omega),
  \label{eqn-appC2}
\end{equation}
and that
the CM noise is much larger than the DM noise:
\begin{equation}
S_{\bar\omega}(\omega)\gg{}S_{\Delta\omega}(\omega).
  \label{eqn-appC3}
\end{equation}

Then, from Eqs.~\eqref{eqn-appC1}, \eqref{eqn-appC2} and
\eqref{eqn-appC3}, the following relation can be shown:
\begin{equation}
 S_{\bar\omega}(\omega)
  \simeq
 S_{\delta\omega_{y,1}}(\omega)
  \simeq
 S_{\delta\omega_{y,2}}(\omega).
\end{equation}

\section{Formalism for heating of radial vibrational modes of two-ion string}
\label{app:4}

Here, we extend the formalism for the heating of a single ion given in
Sec.~\ref{sec:heating} to the case of the radial vibrational modes of a
two-ion string.

We assume that two ions are trapped in a linear trap and 
the $i$th ion ($i=1,2$) is subject to a fluctuating electric
field $\epsilon_i(t)$.
The Hamiltonian in this case is described as:
\begin{equation}
 \hat{H}_2(t)=\hat{H}_{2,0}-q\epsilon_1(t)\hat{y}_1-q\epsilon_2(t)\hat{y}_2,
\end{equation}
where
\begin{eqnarray}
 &&\hat{H}_{2,0}=\sum_{i=1,2}\hat{p}_i^2/2m\\
  &&+\frac{1}{2}m\omega_\mathrm{COM}^2\hat{Q}_\mathrm{COM}^2
  +\frac{1}{2}m\omega_\mathrm{Rock}^2\hat{Q}_\mathrm{Rock}^2,\\
 &&\hat{Q}_\mathrm{COM}=\frac{1}{\sqrt{2}}(\hat{y}_1+\hat{y}_2),\\
 &&\hat{Q}_\mathrm{Rock}=\frac{1}{\sqrt{2}}(\hat{y}_1-\hat{y}_2).
\end{eqnarray}
We define the CM [DM] fluctuating electric field
$\epsilon_\mathrm{CM}(t)$ [$\epsilon_\mathrm{DM}(t)$]
as follows:
\begin{eqnarray}
 \epsilon_\mathrm{CM}(t)&\equiv&\frac{1}{\sqrt{2}}(\epsilon_1(t)+\epsilon_2(t)),\nonumber\\
 \epsilon_\mathrm{DM}(t)&\equiv&\frac{1}{\sqrt{2}}(\epsilon_1(t)-\epsilon_2(t)).\nonumber
\end{eqnarray}
Hence:
\begin{eqnarray}
 \epsilon_1(t)&=&\frac{1}{\sqrt{2}}(\epsilon_\mathrm{CM}(t)+\epsilon_\mathrm{DM}(t)),\nonumber\\
 \epsilon_2(t)&=&\frac{1}{\sqrt{2}}(\epsilon_\mathrm{CM}(t)-\epsilon_\mathrm{DM}(t)).\nonumber
\end{eqnarray}
Then:
\begin{eqnarray}
 \hat{H}_2(t)&=&\hat{H}_{2,0}
  -q\frac{1}{\sqrt{2}}(\epsilon_\mathrm{CM}(t)+\epsilon_\mathrm{DM}(t))\hat{y}_1\nonumber\\
  &&-q\frac{1}{\sqrt{2}}(\epsilon_\mathrm{CM}(t)-\epsilon_\mathrm{DM}(t))\hat{y}_2\nonumber\\
 &=&\hat{H}_{2,0}-q\epsilon_\mathrm{CM}(t)\hat{Q}_\mathrm{COM}-q\epsilon_\mathrm{DM}(t)\hat{Q}_\mathrm{Rock}.\nonumber\\
\end{eqnarray}
This has a form similar to Eq.~\eqref{eqn-H-1}, and the heating rates
are obtained similarly to Eq.~\eqref{eqn-Gamma-0to1} as follows:
\begin{eqnarray}
 \Gamma_{\mathrm{COM},0\to{1}}
  &=&\frac{q^2}{2m\hbar\omega_\mathrm{COM}}S_{1,\mathrm{CM}}(\omega_\mathrm{COM}),
  \\
 \Gamma_{\mathrm{Rock},0\to{1}}
  &=&\frac{q^2}{2m\hbar\omega_\mathrm{Rock}}S_{1,\mathrm{DM}}(\omega_\mathrm{Rock}),
\end{eqnarray} 
where
\begin{eqnarray}
  S_{1,\mathrm{CM}}(\omega)
   &\equiv&\int_{-\infty}^\infty{}d\tau\,{}
   e^{i\omega{}\tau}\langle{}\epsilon_\mathrm{CM}(t)\epsilon_\mathrm{CM}(t+\tau)\rangle,\nonumber\\
  S_{1,\mathrm{DM}}(\omega)
   &\equiv&\int_{-\infty}^\infty{}d\tau\,{}
   e^{i\omega{}\tau}\langle{}\epsilon_\mathrm{DM}(t)\epsilon_\mathrm{DM}(t+\tau)\rangle.\nonumber
\end{eqnarray}

\section{Justification of proportionality relation for noise spectral components}
\label{app:5}

In Sec.~\ref{sec:proportionality},
we considered a simplified situation where the noise
that acts on the trapped ions is dominated by a single noise source
with a particular spectrum, which generates a fluctuating electric
potential with a particular spatial variation. 
Then, 
we assumed the following proportionality
relation for the noise spectral components [Eq.~\eqref{eqn-proportionality1}]:
\begin{eqnarray}
 &&S_{2,\mathrm{CM}}(0):
 S_{2,\mathrm{DM}}(0)\nonumber\\
 &=&
 S_{1,\mathrm{CM}}(\omega_\mathrm{COM}):
 S_{1,\mathrm{DM}}(\omega_\mathrm{Rock}).\nonumber
\end{eqnarray}
In this appendix, we provide the justification for this relation.

We will examine the following two relations in order:
\begin{eqnarray}
 &&S_{1,\mathrm{CM}}(\omega_\mathrm{COM}):
 S_{1,\mathrm{DM}}(\omega_\mathrm{Rock})\nonumber\\
&\approx&
 S_{1,\mathrm{CM}}(0):
 S_{1,\mathrm{DM}}(0),
 \label{eqn-app-5-1}
 \\
 &&S_{1,\mathrm{CM}}(0):S_{1,\mathrm{DM}}(0)\nonumber\\
 &\approx&S_{2,\mathrm{CM}}(0):S_{2,\mathrm{DM}}(0).
  \label{eqn-app-5-2}
\end{eqnarray}

Regarding the relation Eq.~\eqref{eqn-app-5-1}, we consider the
frequency dependence of noise spectral components.  As stated above, we
consider a situation where the noise is dominated by a single source.
In that case, it is reasonable to assume that the spatial profile of the
electric potential is approximately similar for different frequencies.
Then, the ratio of the noise spectral components for two different
spatial geometries does not change considerably when the frequency is
changed, and the following relation holds for arbitrary
$\omega_\mathrm{a}$ and $\omega_\mathrm{b}$:
\begin{eqnarray}
 &&S_{1,\mathrm{CM}}(\omega_\mathrm{a}):
 S_{1,\mathrm{DM}}(\omega_\mathrm{a})\nonumber\\
&\approx&
 S_{1,\mathrm{CM}}(\omega_\mathrm{b}):
 S_{1,\mathrm{DM}}(\omega_\mathrm{b}).
 \label{eqn-app-5-3}
\end{eqnarray}
%
We also consider the fact that the frequency difference between
$\omega_\mathrm{COM}$ and $\omega_\mathrm{Rock}$ ($\approx
2\pi\times200$ kHz in our case) is small compared with their absolute
values ($\approx 2\pi\times3$ MHz). Hence, they can be regarded as
effectively equivalent quantities when compared with the zero frequency
$\omega=0$.  Then, by assuming $\omega_\mathrm{a}=0$ and
$\omega_\mathrm{b}\approx\omega_\mathrm{COM}\approx\omega_\mathrm{Rock}$
in Eq.~ \eqref{eqn-app-5-3}, the relation Eq.~\eqref{eqn-app-5-1} is confirmed.

Regarding the relation Eq.~\eqref{eqn-app-5-2}, we consider the
spatial variation of the fluctuating electric potential.  We take into
account the fact that the distance between the ions ($\sim$10--30~$\mu$m in
our setup) is small compared with other objects in the apparatus
including the trap electrodes (the minimum distance between the ions and
their surface is $\approx0.6$ mm), and hence the electric potential
varies smoothly over the distance.  This fact is reflected, for example,
in the smaller heating rate for the rocking mode due to DM electric-field
noise, compared with that for the COM mode due to CM electric-field
noise.  We show in Sec.~\ref{sec:eval-decay-rate-noise} that the
measured heating rate for the rocking mode is lower than that for the
COM mode by more than one order of magnitude.   Therefore, it is safe to assume
that the local potentials around the two ions are similar, and in that
case it can be shown that:
\begin{eqnarray}
 &&S_{1,\mathrm{Ion1}}(0):S_{2,\mathrm{Ion1}}(0)\nonumber\\
 &\approx&S_{1,\mathrm{Ion2}}(0):S_{2,\mathrm{Ion2}}(0)\nonumber\\
 &\approx&S_{1,\mathrm{CM}}(0):S_{2,\mathrm{CM}}(0)\nonumber\\
 &\approx&S_{1,\mathrm{DM}}(0):S_{2,\mathrm{DM}}(0),
\end{eqnarray} 
where $S_{1,\mathrm{Ion1}}(\omega)$ and
$S_{1,\mathrm{Ion2}}(\omega)$ are the electric-field noise spectra for
the two ions.  
Therefore, the relation Eq.~\eqref{eqn-app-5-2} is confirmed.

Since both Eqs.~\eqref{eqn-app-5-1} and \eqref{eqn-app-5-2} hold, it is
reasonable to assume the proportionality relation in
Eq.~\eqref{eqn-proportionality1}.

\section*{Acknowledgments}
We wish to thank S. Kume and R. Ohira for their contributions to
this study in its early stage. This work was supported by
the Ministry of Education, Culture, Sports, Science and Technology
(MEXT) Quantum Leap Flagship Program (MEXT Q-LEAP) (grant number
JPMXS0118067477) and the Japan Science and Technology Agency Moonshot
Research and Development Program (grant number JPMJMS2063).


\bibliography{main}

@article{cirac1995quantum,
  title = {Quantum computations with cold trapped ions},
  author = {Cirac, Juan I. and Zoller, Peter},
  journal = {Phys. Rev. Lett.},
  volume = {74},
  number = {20},
  pages = {4091--4094},
  year = {1995},
  publisher = {American Physical Society}
}

@article{blatt2008entangled,
  title = {Entangled states of trapped atomic ions},
  author = {Blatt, Rainer and Wineland, David},
  journal = {Nature (London)},
  volume = {453},
  number = {7198},
  pages = {1008--1015},
  year = {2008},
  publisher = {Nature Publishing Group}
}

@article{haffner2008quantum,
  title = {Quantum computing with trapped ions},
  author = {H{\"a}ffner, Hartmut and Roos, Christian F. and Blatt, Rainer},
  journal = {Phys. Rep.},
  volume = {469},
  number = {4},
  pages = {155--203},
  year = {2008},
  publisher = {Elsevier}
}

@article{monroe2013scaling,
  title = {Scaling the ion trap quantum processor},
  author = {Monroe, Christopher and Kim, Jungsang},
  journal = {Science},
  volume = {339},
  number = {6124},
  pages = {1164--1169},
  year = {2013},
  publisher = {American Association for the Advancement of Science}
}

@article{leibfried2003quantum,
  title = {Quantum dynamics of single trapped ions},
  author = {Leibfried, Dietrich and Blatt, Rainer and Monroe, Christopher and Wineland, David},
  journal = {Rev. Mod. Phys.},
  volume = {75},
  number = {1},
  pages = {281--324},
  year = {2003},
  publisher = {American Physical Society}
}

@article{brown2016co,
  title = {Co-designing a scalable quantum computer with trapped atomic ions},
  author = {Brown, Kenneth R. and Kim, Jungsang and Monroe, Christopher},
  journal = {npj Quantum Inf.},
  volume = {2},
  number = {1},
  pages = {1--10},
  year = {2016},
  publisher = {Nature Publishing Group}
}

@article{porras2004effective,
  title = {Effective quantum spin systems with trapped ions},
  author = {Porras, Diego and Cirac, J. Ignacio},
  journal = {Phys. Rev. Lett.},
  volume = {92},
  number = {20},
  pages = {207901},
  year = {2004},
  publisher = {American Physical Society}
}

@article{haze2012observation,
  title = {Observation of phonon hopping in radial vibrational modes of trapped ions},
  author = {Haze, Shinsuke and Tateishi, Yusuke and Noguchi, Atsushi and Toyoda, Kenji and Urabe, Shinji},
  journal = {Phys. Rev. A},
  volume = {85},
  number = {3},
  pages = {031401},
  year = {2012},
  publisher = {American Physical Society}
}

@article{zwerger2003mott,
  title = {Mott--{H}ubbard transition of cold atoms in optical lattices},
  author = {Zwerger, Wilhelm},
  journal = {J. Opt. B},
  volume = {5},
  number = {2},
  pages = {S9--S16},
  year = {2003},
  publisher = {IOP Publishing}
}

@article{leggett1987dynamics,
  title = {Dynamics of the dissipative two-state system},
  author = {Leggett, Anthony J. and Chakravarty, Sudip and Dorsey, Alan T. and Fisher, Matthew P. A. and Garg, Anupam and Zwerger, Wilhelm},
  journal = {Rev. Mod. Phys.},
  volume = {59},
  number = {1},
  pages = {1--85},
  year = {1987},
  publisher = {American Physical Society}
}

@article{holstein1959studies,
  title = {Studies of polaron motion: {P}art {I}. {T}he molecular-crystal model},
  author = {Holstein, T.},
  journal = {Ann. Phys.},
  volume = {8},
  number = {3},
  pages = {325--342},
  year = {1959},
  publisher = {Elsevier}
}

@book{fehske2007numerical,
  author = {Fehske, Holger and Trugman, Stuart A.},
  title = {Numerical solution of the {H}olstein polaron problem},
  booktitle = {Polarons in Advanced Materials},
  series = {Springer Series in Materials Science},
  volume = {103},
  editor = {Alexandrov, A.~S.},
  publisher = {Springer Netherlands},
  pages = {393--461},
  year = {2007}
}

@article{schmidt2003realization,
  title = {Realization of the {C}irac--{Z}oller controlled-{NOT} quantum gate},
  author = {Schmidt-Kaler, Ferdinand and H{\"a}ffner, Hartmut and Riebe, Mark and Gulde, Stephan and Lancaster, Gavin P. T. and Deuschle, Thomas and Becher, Christoph and Roos, Christian F. and Eschner, J{\"u}rgen and Blatt, Rainer},
  journal = {Nature (London)},
  volume = {422},
  number = {6930},
  pages = {408--411},
  year = {2003},
  publisher = {Nature Publishing Group}
}

@article{greiner2002quantum,
  title = {Quantum phase transition from a superfluid to a {M}ott insulator in a gas of ultracold atoms},
  author = {Greiner, M. and Mandel, O. and Esslinger, T. and H{\"a}nsch, T. W. and Bloch, I.},
  journal = {Nature (London)},
  volume = {415},
  number = {6867},
  pages = {39--44},
  year = {2002},
  publisher = {Nature Publishing Group}
}

@book{reif2009fundamentals,
  title = {Fundamentals of Statistical and Thermal Physics},
  author = {Reif, Frederick},
  year = {2009},
  publisher = {Waveland Press},
  isbn = {978-1-57766-612-7}
}

@article{johnson2016active,
  title = {Active stabilization of ion trap radiofrequency potentials},
  author = {Johnson, K. G. and Wong-Campos, J. D. and Restelli, A. and Landsman, K. A. and Neyenhuis, B. and Mizrahi, J. and Monroe, C.},
  journal = {Rev. Sci. Instrum.},
  volume = {87},
  number = {5},
  pages = {053110},
  year = {2016},
  publisher = {AIP Publishing}
}

@article{porras2004bose,
  title = {Bose-{E}instein condensation and strong-correlation behavior of phonons in ion traps},
  author = {Porras, D. and Cirac, J. I.},
  journal = {Phys. Rev. Lett.},
  volume = {93},
  number = {26},
  pages = {263602},
  year = {2004},
  publisher = {American Physical Society}
}

@article{roos2008nonlinear,
  title = {Nonlinear coupling of continuous variables at the single quantum level},
  author = {Roos, C. F. and Monz, T. and Kim, K. and Riebe, M. and H{\"a}ffner, H. and James, D. F. V. and Blatt, R.},
  journal = {Phys. Rev. A},
  volume = {77},
  number = {4},
  pages = {040302},
  year = {2008},
  publisher = {American Physical Society}
}

@article{james1998quantum,
  title = {Quantum dynamics of cold trapped ions with application to quantum computation},
  author = {James, D. F. V.},
  journal = {Appl. Phys. B},
  volume = {66},
  pages = {181},
  year = {1998}
}

@article{cui2022evaluation,
  title = {Evaluation of the systematic shifts of a $^{40}${C}a$^+$--$^{27}${A}l$^+$ optical clock},
  author = {Cui, K. and Chao, S. and Sun, C. and Wang, S. and Zhang, P. and Wei, Y. and Yuan, J. and Cao, J. and Shu, H. and Huang, X.},
  journal = {Eur. Phys. J. D},
  volume = {76},
  number = {8},
  pages = {140},
  year = {2022},
  publisher = {Springer}
}

@article{Huang2014Evaluation,
  title = {Evaluation of the systematic shifts and absolute frequency measurement of a single {C}a$^+$ ion frequency standard},
  author = {Huang, Y. and Liu, P. and Bian, W. and Guan, H. and Gao, K.},
  journal = {Appl. Phys. B},
  volume = {114},
  pages = {189--201},
  year = {2014},
  publisher = {Springer}
}

@article{nie2009theory,
  title = {Theory of cross phase modulation for the vibrational modes of trapped ions},
  author = {Nie, X. R. and Roos, C. F. and James, D. F. V.},
  journal = {Phys. Lett. A},
  volume = {373},
  number = {4},
  pages = {422--425},
  year = {2009},
  publisher = {Elsevier}
}

@article{Brown2011,
   author = {Brown, K. R. and Ospelkaus, C. and Colombe, Y. and Wilson, A. C. and Leibfried, D. and Wineland, D. J.},
   title = {Coupled quantized mechanical oscillators},
   journal = {Nature (London)},
   volume = {471},
   number = {7337},
   pages = {196-199},
   ISSN = {0028-0836},
   DOI = {Doi 10.1038/Nature09721},
   url = {<Go to ISI>://000288170200032
http://www.nature.com/nature/journal/v471/n7337/pdf/nature09721.pdf},
   year = {2011},
   type = {Journal Article}
}

@article{Harlander2011,
   author = {Harlander, M. and Lechner, R. and Brownnutt, M. and Blatt, R. and Hansel, W.},
   title = {Trapped-ion antennae for the transmission of quantum information},
   journal = {Nature (London)},
   volume = {471},
   number = {7337},
   pages = {200-203},
   ISSN = {0028-0836},
   DOI = {10.1038/nature09800},
   url = {<Go to ISI>://000288170200033},
   year = {2011},
   type = {Journal Article}
}

@article{Wilson2014,
   author = {Wilson, A. C. and Colombe, Y. and Brown, K. R. and Knill, E. and Leibfried, D. and Wineland, D. J.},
   title = {Tunable spin–spin interactions and entanglement of ions in separate potential wells},
   journal = {Nature (London)},
   volume = {512},
   number = {7512},
   pages = {57-60},
   ISSN = {1476-4687},
   DOI = {10.1038/nature13565},
   url = {https://doi.org/10.1038/nature13565},
   year = {2014},
   type = {Journal Article}
}

@article{Hakelberg2019,
   author = {Hakelberg, F. and Kiefer, P. and Wittemer, M. and Warring, U. and Schaetz, T.},
   title = {Interference in a Prototype of a Two-Dimensional Ion Trap Array Quantum Simulator},
   journal = {Phys. Rev. Lett.},
   volume = {123},
   number = {10},
   pages = {100504},
   ISSN = {0031-9007},
   DOI = {Artn 100504
10.1103/Physrevlett.123.100504},
   url = {<Go to ISI>://000483944300001},
   year = {2019},
   type = {Journal Article}
}

@article{Kiefer2019,
   author = {Kiefer, P. and Hakelberg, F. and Wittemer, M. and Bermudez, A. and Porras, D. and Warring, U. and Schaetz, T.},
   title = {Floquet-Engineered Vibrational Dynamics in a Two-Dimensional Array of Trapped Ions},
   journal = {Phys. Rev. Lett.},
   volume = {123},
   number = {21},
   pages = {213605},
   ISSN = {0031-9007},
   DOI = {Artn 213605
10.1103/Physrevlett.123.213605},
   url = {<Go to ISI>://000498063600002},
   year = {2019},
   type = {Journal Article}
}

@article{Hou2024,
   author = {Hou, Pan-Yu and Wu, Jenny J. and Erickson, Stephen D. and Cole, Daniel C. and Zarantonello, Giorgio and Brandt, Adam D. and Geller, Shawn and Kwiatkowski, Alex and Glancy, Scott and Knill, Emanuel and Wilson, Andrew C. and Slichter, Daniel H. and Leibfried, Dietrich},
   title = {Coherent coupling and non-destructive measurement of trapped-ion mechanical oscillators},
   journal = {Nat. Phys.},
   volume = {20},
   pages = {1636},
   ISSN = {1745-2481},
   DOI = {10.1038/s41567-024-02585-y},
   url = {https://doi.org/10.1038/s41567-024-02585-y},
   year = {2024},
   type = {Journal Article}
}

@article{Chen2023,
   author = {Chen, Wentao and Lu, Yao and Zhang, Shuaining and Zhang, Kuan and Huang, Guanhao and Qiao, Mu and Su, Xiaolu and Zhang, Jialiang and Zhang, Jing-Ning and Banchi, Leonardo and Kim, M. S. and Kim, Kihwan},
   title = {Scalable and programmable phononic network with trapped ions},
   journal = {Nat. Phys.},
   volume = {19},
   number = {6},
   pages = {877-883},
   ISSN = {1745-2481},
   DOI = {10.1038/s41567-023-01952-5},
   url = {https://doi.org/10.1038/s41567-023-01952-5
https://www.nature.com/articles/s41567-023-01952-5.pdf},
   year = {2023},
   type = {Journal Article}
}

@article{Gan2020,
   author = {Gan, H. C. J. and Maslennikov, Gleb and Tseng, Ko-Wei and Nguyen, Chihuan and Matsukevich, Dzmitry},
   title = {Hybrid Quantum Computing with Conditional Beam Splitter Gate in Trapped Ion System},
   journal = {Phys. Rev. Lett.},
   volume = {124},
   number = {17},
   pages = {170502},
   DOI = {10.1103/PhysRevLett.124.170502},
   url = {https://link.aps.org/doi/10.1103/PhysRevLett.124.170502},
   year = {2020},
   type = {Journal Article}
}

@article{Zhu2006,
   author = {Zhu, Shi-Liang and Monroe, C. and Duan, L. M.},
   title = {Trapped Ion Quantum Computation with Transverse Phonon Modes},
   journal = {Phys. Rev. Lett.},
   volume = {97},
   number = {5},
   pages = {050505},
   url = {http://link.aps.org/abstract/PRL/v97/e050505 },
   year = {2006},
   type = {Journal Article}
}

@article{Tamura2020,
   author = {Tamura, M. and Mukaiyama, T. and Toyoda, K.},
   title = {Quantum Walks of a Phonon in Trapped Ions},
   journal = {Phys. Rev. Lett.},
   volume = {124},
   number = {20},
   pages = {200501},
   ISSN = {0031-9007},
   DOI = {ARTN 200501
10.1103/PhysRevLett.124.200501},
   url = {<Go to ISI>://WOS:000533812900001
https://journals.aps.org/prl/pdf/10.1103/PhysRevLett.124.200501},
   year = {2020},
   type = {Journal Article}
}

@article{Toyoda2015,
   author = {Toyoda, Kenji and Hiji, Ryoto and Noguchi, Atsushi and Urabe, Shinji},
   title = {{H}ong–{O}u–{M}andel interference of two phonons in trapped ions},
   journal = {Nature (London)},
   volume = {527},
   number = {7576},
   pages = {74-77},
   ISSN = {0028-0836},
   DOI = {10.1038/nature15735},
   url = {http://dx.doi.org/10.1038/nature15735},
   year = {2015},
   type = {Journal Article}
}

@article{Debnath2018,
   author = {Debnath, S. and Linke, N. M. and Wang, S. T. and Figgatt, C. and Landsman, K. A. and Duan, L. M. and Monroe, C.},
   title = {Observation of Hopping and Blockade of Bosons in a Trapped Ion Spin Chain},
   journal = {Phys. Rev. Lett.},
   volume = {120},
   number = {7},
   pages = {073001},
   ISSN = {0031-9007},
   DOI = {Artn 073001
10.1103/Physrevlett.120.073001},
   url = {<Go to ISI>://000424750200005},
   year = {2018},
   type = {Journal Article}
}

@article{Li2022,
   author = {Li, B. W. and Mei, Q. X. and Wu, Y. K. and Cai, M. L. and Wang, Y. and Yao, L. and Zhou, Z. C. and Duan, L. M.},
   title = {Observation of Non-{M}arkovian Spin Dynamics in a {J}aynes-{C}ummings-{H}ubbard Model Using a Trapped-Ion Quantum Simulator},
   journal = {Phys. Rev. Lett.},
   volume = {129},
   number = {14},
   pages = {140501},
   DOI = {10.1103/PhysRevLett.129.140501},
   url = {https://link.aps.org/doi/10.1103/PhysRevLett.129.140501},
   year = {2022},
   type = {Journal Article}
}

@article{Mei2022,
   author = {Mei, Q. X. and Li, B. W. and Wu, Y. K. and Cai, M. L. and Wang, Y. and Yao, L. and Zhou, Z. C. and Duan, L. M.},
   title = {Experimental Realization of the {R}abi-{H}ubbard Model with Trapped Ions},
   journal = {Phys. Rev. Lett.},
   volume = {128},
   number = {16},
   pages = {160504},
   DOI = {10.1103/PhysRevLett.128.160504},
   url = {https://link.aps.org/doi/10.1103/PhysRevLett.128.160504
https://journals.aps.org/prl/abstract/10.1103/PhysRevLett.128.160504},
   year = {2022},
   type = {Journal Article}
}

@article{Toyoda2013,
   author = {Toyoda, K. and Matsuno, Y. and Noguchi, A. and Haze, S. and Urabe, S.},
   title = {Experimental Realization of a Quantum Phase Transition of Polaritonic Excitations},
   journal = {Phys. Rev. Lett.},
   volume = {111},
   number = {16},
   pages = {160501},
   ISSN = {0031-9007},
   DOI = {Artn 160501
Doi 10.1103/Physrevlett.111.160501},
   url = {<Go to ISI>://000326119300001
http://prl.aps.org/pdf/PRL/v111/i16/e160501},
   year = {2013},
   type = {Journal Article}
}

@article{Ivanov2009,
   author = {Ivanov, P. A. and Ivanov, S. S. and Vitanov, N. V. and Mering, A. and Fleischhauer, M. and Singer, K.},
   title = {Simulation of a quantum phase transition of polaritons with trapped ions},
   journal = {Phys. Rev. A},
   volume = {80},
   number = {6},
   pages = {060301(R)},
   ISSN = {1050-2947},
   DOI = {Artn 060301
Doi 10.1103/Physreva.80.060301},
   url = {<Go to ISI>://000273233800002
http://pra.aps.org/abstract/PRA/v80/i6/e060301},
   year = {2009},
   type = {Journal Article}
}

@article{Shen2014,
   author = {Shen, C. and Zhang, Z. and Duan, L. M.},
   title = {Scalable Implementation of Boson Sampling with Trapped Ions},
   journal = {Phys. Rev. Lett.},
   volume = {112},
   number = {5},
   pages = {050504},
   ISSN = {0031-9007},
   DOI = {Artn 050504
Doi 10.1103/Physrevlett.112.050504},
   url = {<Go to ISI>://000331950100003},
   year = {2014},
   type = {Journal Article}
}

@article{Chen2021,
   author = {Chen, W. T. and Gan, J. and Zhang, J. N. and Matuskevich, D. and Kim, K.},
   title = {Quantum computation and simulation with vibrational modes of trapped ions},
   journal = {Chin. Phys. B},
   volume = {30},
   number = {6},
   pages = {060311},
   ISSN = {1674-1056},
   DOI = {ARTN 060311
10.1088/1674-1056/ac01e3},
   url = {<Go to ISI>://WOS:000668706700001},
   year = {2021},
   type = {Journal Article}
}

@article{savard1997laser,
  title={Laser-noise-induced heating in far-off resonance optical traps},
  author={Savard, TA and O’hara, KM and Thomas, JE},
  journal={Phys. Rev. A},
  volume={56},
  number={2},
  pages={R1095},
  year={1997},
  publisher={APS}
}

@article{turchette2000heating,
  title={Heating of trapped ions from the quantum ground state},
  author={Turchette, Q. A. and Kielpinski and King, B. E. and Leibfried, D. and Meekhof, D. M. and Myatt, C. J. and Rowe, M. A. and Sackett, C. A. and Wood, C. S. and Itano, W. M. and Monroe, C. and Wineland, D. J.},
  journal={Phys. Rev. A},
  volume={61},
  number={6},
  pages={063418},
  year={2000},
  publisher={APS}
}

@article{bergli2009decoherence,
  title={Decoherence in qubits due to low-frequency noise},
  author={Bergli, J and Galperin, Yu M and Altshuler, BL},
  journal={New J. Phys.},
  volume={11},
  number={2},
  pages={025002},
  year={2009},
  publisher={IOP Publishing}
}

@article{Deslauriers2006,
   author = {Deslauriers, L. and Olmschenk, S. and Stick, D. and Hensinger, W. K. and Sterk, J. and Monroe, C.},
   title = {Scaling and Suppression of Anomalous Heating in Ion Traps},
   journal = {Phys. Rev. Lett.},
   volume = {97},
   number = {10},
   pages = {103007},
   url = {http://link.aps.org/abstract/PRL/v97/e103007 },
   year = {2006},
   type = {Journal Article}
}

@article{Degen2017,
   author = {Degen, C. L. and Reinhard, F. and Cappellaro, P.},
   title = {Quantum sensing},
   journal = {Rev. Mod. Phys.},
   volume = {89},
   number = {3},
   pages = {035002},
   ISSN = {0034-6861},
   DOI = {ARTN 035002
10.1103/RevModPhys.89.035002},
   url = {<Go to ISI>://WOS:000406358200001},
   year = {2017},
   type = {Journal Article}
}

@misc{Preskill2006,
  author       = {John Preskill},
  title        = {Notes on noise},
  howpublished = {\url{https://www.preskill.caltech.edu/papers/decoherence_notes.pdf}},
  note         = {Updated 2 December 2006, accessed 28 Jan 2026},
  year         = {2006}
}

@article{Roos2000,
   author = {Roos, C. F. and Leibfried, D. and Mundt, A. and Schmidt-Kaler, F. and Eschner, J. and Blatt, R.},
   title = {Experimental demonstration of ground state laser cooling with electromagnetically induced transparency},
   journal = {Phys. Rev. Lett.},
   volume = {85},
   number = {26},
   pages = {5547-5550},
   ISSN = {0031-9007},
   DOI = {DOI 10.1103/PhysRevLett.85.5547},
   url = {<Go to ISI>://WOS:000166132500017},
   year = {2000},
   type = {Journal Article}
}

\end{document}